\documentclass{iopjournal}

\usepackage{graphicx}
\usepackage{dcolumn}
\usepackage{amsmath}
\usepackage{bm}
\usepackage{hyperref}
\usepackage{color}
\usepackage{float}
\usepackage[
    style=numeric-comp,
    maxcitenames=2,
    maxbibnames=10,
    sorting=none,
    backend=bibtex
]{biblatex}

\definecolor{mypurple}{rgb}{0.5, 0, 0.85}
\definecolor{Hbetta}{rgb}{0,0.92,1}
\definecolor{myblue}{rgb}{0, 0.2, 0.85}
\definecolor{cadmiumgreen}{rgb}{0.0, 0.42, 0.24}
\definecolor{gold}{rgb}{0.7176, 0.5843, 0.0431}
\definecolor{org}{RGB}{33, 47, 61}

\newcommand{\msun}{\ensuremath{M_\odot}}

\begin{document}

\title{Universal relations applied to proto-neutron star generated gravitational waves from three-dimensional core collapse supernova simulations} 

\author{R. Daniel Murphy$^{1,*}$\orcid{0000-0003-4459-2557}, Anthony Mezzacappa$^{1}$\orcid{0000-0001-9816-9741}, Colter J. Richardson$^{1}$\orcid{0000-0003-1866-7965}, Pedro Marronetti$^{2}$\orcid{0000-0003-3070-0625}, Eric J. Lentz$^{1,3}$\orcid{0000-0002-5231-0532}, Ryan E. Landfield$^{4}$}

\affil{$^1$Department of Physics and Astronomy, University of Tennessee, Knoxville, TN 37996, USA}\\
\affil{$^2$Physics Division, National Science Foundation, Alexandria, Virginia 22314, USA}\\
\affil{$^3$Physics Division, Oak Ridge National Laboratory, P.O. Box 2008, Oak Ridge, Tennessee 37831-6354, USA}\\
\affil{$^4$National Center for Computational Sciences, Oak Ridge National Laboratory, P.O. Box 2008, Oak Ridge, Tennessee 37831-6164, USA}\\
\affil{$^*$Author to whom any correspondence should be addressed.}

\email{rmurph16@utk.edu}

\date{\today}


\begin{abstract}
Using asteroseismology techniques, several relations have been developed that relate the quasi-normal, non-radial oscillation mode frequencies of the proto-neutron star (PNS) to the high frequency component of core collapse supernova (CCSN) generated gravitational waves predicted from simulation. These relations are universal in the sense that they are parameterized entirely by PNS properties, e.g., mean density or surface gravity, and are independent of both progenitor properties, e.g., zero age man sequence (ZAMS) mass or metallicity, and the physics included in CCSN simulations, e.g., nuclear equation of state (EOS). In this work, we apply several externally developed universal relations to PNS evolution data---specifically the mass, \textit{M}, and radius, \textit{R},---generated from both two- and three-dimensional CCSN simulations and compare the resulting PNS oscillation frequencies predicted by the universal relations to the peak gravitational wave frequencies computed directly from the simulation-produced spectrogram. Additionally, we use the gravitational wave spectrogram peak frequencies as input to the universal relations and compare the predicted PNS properties from each relation to the true PNS properties as determined by the simulations. In this way, we show what these universal relations would predict for the PNS properties and their evolution from a real CCSN gravitational wave detection in the best case scenario, i.e., no detector noise. Our results indicate that caution must be exercised when using these universal relations, particularly when interpreting their predictions for PNS evolution from a gravitational wave detection, and that, given the extent to which we do see agreement between asteroseismological predictions and simulation outcomes, further development of universal relations would be beneficial. 
\end{abstract}

\section{Introduction}
\label{sec:introduction}

The detection of gravitational waves from a core collapse supernova (CCSN) will be a breakthrough in modern astrophysics. Among the myriad scientific outcomes expected from such a detection is the direct observation of the birth and early evolution of the proto-neutron star (PNS) that exists for a short time in the densest regions of the CCSN, ultimately either cooling to become a neutron star or further collapsing to a black hole. From CCSN simulations \textit{virtual} gravitational wave detections can be constructed, and the expected gravitational wave signal from a CCSN will have contributions broadly separable into low frequency and high frequency components starting $\sim$100 ms after core bounce. The low frequency signal will have contributions from both the standing accretion shock instability and the stochastic motion behind the shock due to turbulent neutrino driven convection \cite{KuKoTa16,KuKoHa17, PoMuHe21,MeMaLa23}. There will be an additional low frequency component to the signal from the anisotropic emission of neutrinos and the asymmetric explosion that will eventually lead to gravitational wave memory \cite{VaBu20,RiZaAn22,RiAnMe24}. The high frequency component of the gravitational wave signal will also have stochastic contributions due to matter impinging on the PNS, but it will be dominated by the oscillations of the PNS that rise in frequency as the PNS contracts \cite{BuLiDe06,MaJaMu09,CeDeAl13,MuJaMa13,SoTa16,AnMuMu17,KaKuTa18,RaMoBu19,PoMu19,MeMaLa20,VaBuWa23}. 

Over the first $\mathcal{O}(1s)$ of evolution of the PNS, the detailed dynamics of the newly formed PNS are dependent on the nuclear equation of state (EOS) and strong interaction physics therein, the transport of neutrinos and weak interaction physics therein, which deleptonizes the core and effectively ``cools'' the PNS, the hydrodynamics of accretion depositing matter onto the surface and ``ringing'' the PNS, and magneto-rotational effects driven by the rotation of the PNS, magnetic fields present, and their evolution. At the same time, it is possible to use methods similar to traditional asteroseismology to investigate the quasi-normal modes of oscillation of the PNS, and it has been shown that the quadrupolar oscillation mode frequencies of the PNS predicted from these analyses align well, starting $\sim$ 100 ms after bounce, with the high frequency gravitational wave signal from simulation \cite{ToCePa18,ToCePa19,MoRaBu18,SoKuTa19,West20,ZhAnOc24,MuMeLe25}.  Despite the complexity of the physics involved, there is growing consensus that the frequency evolution of the oscillation modes of the PNS may be parameterized by the surface gravity of the PNS, mean density of the PNS, or some combination of the two. In this way, so-called ``universal relations'' may exist that connect the evolution of the high frequency component of the CCSN gravitational wave signal directly to the surface gravity or mean density evolution of the PNS independently of currently unknown microphysics or undetermined progenitor properties.
 
In this work, we use data from the \textsc{Chimera} CCSN simulation code \cite{BrBlHi20} to examine the predictions from the universal relations from seven different studies: Torres-Forné et al. 2021 \cite{ToCeMu21}, Sotani and Sumiyoshi 2019 \cite{SoSu19}, Sotani et al. 2021 \cite{SoTaTo21}, Sotani et al. 2025 \cite{SoMuTa25}, Mori et al. 2023 \cite{MoSuTa23}, Rodriguez et al. 2023 \cite{RoRaCh23}, and Bizouard et al. 2021 \cite{BiMaTo21}. These universal relations were derived using differing PNS oscillation analyses on data produced by several different CCSN simulation codes using a wide range of progenitors, differing microphysics, and differing dimensionality. The article is organized as follows: Section \ref{sec:models_and_methods} describes the \textsc{Chimera} models and the data, it contains Section \ref{subsec: UR}, which provides a brief overview of the universal relations we considered, and Section \ref{subsec: applying UR}, which details the construction of our spectrograms and how we compare universal relation predictions to \textsc{Chimera} data. Section \ref{sec:results} discusses the results of both using universal relations to predict PNS oscillation frequencies and using simulation-produced gravitational wave frequencies with the universal relations to predict PNS properties. We end the article with a discussion of our findings and a caution on the uncritical use of universal relations in Section \ref{sec:conclusion}.

\section{Models and Methods}
\label{sec:models_and_methods}
We examined the gravitational wave emission from four models using the \textsc{Chimera} CCSN simulation code \cite{BrBlHi20}. The \textsc{Chimera} code is a multiphysics code that combines Newtonian hydrodynamics with multipole self-gravity that includes general relativistic corrections on the monopole term (\cite{MaDiJa06} case A, henceforth referred to as pseudo-Newtonian gravity), multigroup flux-limited diffusion neutrino transport in the ``ray-by-ray-plus'' approximation \cite{BuRaJa06}, and a nuclear reaction network \cite{HiMe06,TrHi13} to simulate CCSN. Complete details on the \textsc{Chimera} code, including differences between previous versions, can be found in \cite{BrBlHi20}. 

The first set of models consisted of two, three-dimensional simulations from collapse through explosion of a solar-metallicity, non-rotating, 15 \msun\ progenitor from \cite{WoHe07} and a zero-metallicity, non-rotating, 25 \msun\ progenitor from \cite{HeWo10}. These models are a subset of the \textsc{Chimera} D-series simulations described in \cite{MeMaLa23}, with the former model denoted as D15 and the later as D25. The gravitational wave data sourced from matter in these simulations are available at \cite{OLCF_GW_M}, and examined in detail for D15 and D25 in \cite{MuMeLe25}. The EOS used in the D-series is the LSBCK EOS, a combination of the EOS of \cite{LaSw91} with a nuclear incompressibility of $K=220$ MeV at densities above $10^{11}$ g cm$^{-3}$ and the EOS \cite{BaCoKa85} for densities below that.

The second set of \textsc{Chimera} models consisted of two, two-dimensional simulations from the collapse through explosion of the same 15 \msun\ progenitor as D15. Each two-dimensional model used a different EOS. The first used the same EOS as the D-series models, and the second used the SFHo EOS of \cite{StHeFi13}. These models are a subset of the \textsc{Chimera} E-series simulations described in \cite{Landfield_thesis, MuCaMe24}, with the former denoted as E15-LSBCK and the latter as E15-SFHo.

\subsection{Universal Relations Considered}
\label{subsec: UR}

\begin{table}
\caption{Parameters for universal relations using Eq. \ref{eq:UR_base}, $f(x)=c_1+c_2\mathrm{ln}(x)+c_3x+c_4x^2+c_5x^3$. The units of $f$ are KHz. The PNS mass $M$ and radius $R$ are determined using the $10^{11}$ g cm$^{-3}$ density contour as the PNS surface and have units of \msun\ and km, respectively. Quantities determined using the $10^{10}$ g cm$^{-3}$ density contour as the PNS surface are denoted as $M_{10}$ and $R_{10}$. Normalized quantities are denoted as $M^*$ and $R^*$ where $M^*=M/1.4$ and $R^*=R/10$.}
\centering
\begin{tabular}{l c c c c c}
\hline
Reference & $x$ & $c_1$ & $c_2$ & $c_3$ & $c_4$ \\
\hline
Torres-Forné et al. 2021 - ${}^2g_1$ mode \cite{ToCeMu21} & $M_{10}/R_{10}^{2}$ & 0 & 0 & 867 & -51.9$\times10^3$ \\
Torres-Forné et al. 2021 - ${}^2g_2$ mode \cite{ToCeMu21} & $M_{10}/R_{10}^{2}$ & 0 & 0 & 588 & -86.2$\times10^3$ \\
Sotani et al. 2019 \cite{SoSu19} & $\sqrt{M^*/R^*{}^3}$ & 0.9733 & 0 & -2.7171 & -13.7809 \\
Sotani et al. 2021 \cite{SoTaTo21} & $\sqrt{M^*/R^*{}^3}$ & -1.410 & -0.443 & 9.337 & -6.714 \\
Sotani et al. 2021 \cite{SoTaTo21} & $1000\ M/R^{2}$ & -0.0752 & -0.2600 & 0.7446 & 0.0600 \\
Sotani et al. 2025 \cite{SoMuTa25} & $\sqrt{M^*/R^*{}^3}$ & 0.0082 & 0 & 4.5908 & -2.6821\\
Mori et al. 2023 \cite{MoSuTa23} & $\sqrt{M/R^{3}}$ & 3.340 & 0.5303 & 3.399 & 417.6 \\
Mori et al. 2023 \cite{MoSuTa23} & $M/R^{2}$ & 3.264 & 0.3929 & 31.23 & 1962 \\
Mori et al. 2023 \cite{MoSuTa23} & $M/R$ & 5.279 & 1.258 & -19.27 & 128.0 \\
Rodriguez et al. 2023 - ${}^2f$ mode \cite{RoRaCh23} & $\sqrt{M_{10}/R_{10}^{3}}$ & -0.050 & 0 & 174.3 & -2590 \\
Rodriguez et al. 2023 - ${}^2p_1$ mode  \cite{RoRaCh23} & $\sqrt{M_{10}/R_{10}^{3}}$ & 0.067 & 0 & 184.6 & 2194 \\
\hline
\end{tabular}
\label{tab1}
\end{table}

We consider universal relations of the form
\begin{equation}
    f(x)=c_1+c_2\mathrm{ln}(x)+c_3x+c_4x^2+c_5x^3,\label{eq:UR_base}
\end{equation}
where $f$ is the PNS oscillation frequency in kHz, $x$ is the PNS property (e.g., average density), and $c_n$ are constants shown in Table \ref{tab1}. The units of each constant are such that each term in Eq. \ref{eq:UR_base} has units of kHz. Note that all universal relations we considered in this work had $c_5=0$ except for the ${}^2g_2$ mode of Torres-Forné et al. 2021 \cite{ToCeMu21}, for which $c_5=4.67\times10^6$. The general process used to derive each universal relation we consider is as follows: A CCSN simulation computes the hydrodynamic properties of the PNS. These properties are either spherically averaged, if necessary, or used to solve the Tolman-Oppenheimer-Volkoff equations to provide a spherically symmetric PNS structure that is treated as a background. Upon this background, a linear perturbative analysis is performed to determine the quasi-normal, non-radial oscillation modes of the PNS. Some collection of PNS oscillation frequencies that agree well with the virtual gravitational wave signal generated from simulation data were then used to determine the parameters $c_n$ in Eq. \eqref{eq:UR_base}. We note that the classification of these oscillation modes is non-trivial, with differing classifications used by different groups. Three such classifications are explored in \cite{RoRaCh23}, and \cite{ToCePa19} propose an eigenfunction matching procedure to classify modes. In this work, we abstain from commenting on the classification of modes in lieu of investigating the PNS properties. We make no attempt to unify the classification schemes used for each universal relation, and, when necessary, refer to the oscillation modes by the classifications used within each reference.

To explain the differing origins of the universal relations considered in this work, below we provide an overview of the models used to produce them. In particular, for the CCSN simulations used to produce the PNS models we note the spatial dimensionality, EOS used, and treatment of gravity (i.e., relativistic or pseudo-Newtonian). For the linear PNS analysis, we note the treatment of gravity, including if the relativistic Cowling approximation \cite{McvHSc83} was used, and the location of the boundary of the analysis. Following traditional asteroseismology, one choice of boundary is the surface of the PNS where the Lagrangian pressure perturbation should vanish. This can be further complicated by the varying definitions of the surface of the PNS, but in general this is determined by some spherically averaged density contour. Alternatively, to account for the potential effects of the active matter motion that continues outside the PNS surface even after the onset of explosion, the average shock radius can be chosen as the outer boundary where the radial displacement must vanish. This does not represent the totality of the differences present between the origins of the universal relations we consider here, but they are differences, with the exception of EOS, that have been shown to affect the PNS oscillations frequencies determined by linear PNS analysis \cite{SoTaTo19,West20,SoMuTa24,SoMuTa25}. We summarize these differences for each relation considered in Table \ref{tab2}.

\begin{table}
\caption{Summary of select characteristics from each reference that derived a universal relation. We list in each column, from left to right, the reference for universal relation(s), the number of models used, the dimensionality of the CCSN simulations that produced PNS background data, the treatment of gravity in the CCSN simulations (Newtonian, Pseudo-Newt., or GR), the treatment of gravity in the linear PNS analysis, if the relativistic Cowling approximation was used, and the density contour used to identify the surface of the PNS in g cm$^{-3}$.}
\centering
\begin{tabular}{l c c c c c c }
\hline
Reference & \# Models & CCSN & CCSN & PNS & Cowling & PNS \\
 & & dim. & gravity & gravity & approx. & surface \\
\hline
Torres-Forné et al. 2021 \cite{ToCeMu21} & 25 & 1D & GR, & GR & No & $10^{10}$ \\
 & & & Newtonian, & & & \\
 & & & Pseudo-Newt. & & & \\
Sotani et al. 2019 \cite{SoSu19} & 5 & 1D & GR & GR & Yes & $10^{11}$ \\
Sotani et al. 2021 \cite{SoTaTo21} & 5 & 2D & Pseudo-Newt. & GR & Yes & $10^{11}$ \\
Sotani et al. 2025 \cite{SoMuTa25} & 2 & 2D & GR & GR & No & $10^{11}$ \\
Mori et al. 2023 \cite{MoSuTa23} & 1 & 1D & GR & GR & No & $10^{11}$ \\
Rodriguez et al. 2023 \cite{RoRaCh23} & 7 & 3D & Pseudo-Newt. & GR & Yes & $10^{10}$ \\
\hline
\end{tabular}
\label{tab2}
\end{table}

In Torres-Forné et al. 2021 \cite{ToCeMu21}, the relations first derived in Torres-Forné et al. 2019 \cite{ToCeOb19} are updated and based on 25 one-dimensional CCSN simulations using the \small{AENUS-ALCAR} \cite{Obergaulinger_thesis,JuObJa15} and \small{CoCoNut} \cite{DiNoFo05} codes. The pre-supernova models used were from \cite{WoHeWe02} and spanned a mass range of 11 \msun\ to 75 \msun\ with solar metallicity and one 20 \msun\ model with $10^{-4}$ solar metallicity. The treatment of gravity in the CCSN simulations across models varied from Newtonian, pseudo-Newtonian, and a general relativistic (GR) treatment using the extended conformal flatness condition (XCFC) approximation \cite{CoCeDi09}. The linear PNS analysis is performed using the GREAT code \cite{ToCePa18,ToCePa19}, where gravity was treated using the conformal flatness condition (CFC) approximation \cite{WiMaMa96,Isenberg08} and metric perturbations are included; i.e., the relativistic Cowling approximation was not used. Six different EOS were considered across models: LS220 \cite{LaSw91}, GShen-NL3 \cite{ShHoTe11}, HShen \cite{ShToOy11}, SFHo \cite{StHeFi13}, BHB-$\Lambda$ \cite{BaHeBa14}, and HShen-$\Lambda$ \cite{ShToOy11}. Universal relations were determined for several PNS oscillation modes using as the outer boundary for the linear PNS analysis either the shock radius or the PNS surface defined as the spherically averaged $10^{10}$ g cm$^{-3}$ density contour.

In Sotani and Sumiyoshi 2019 \cite{SoSu19}, five spherically symmetric CCSN models are considered using a relativistic neutrino radiation hydrodynamics code \cite{Yamada97,YaJaSu99,SuYaSu05}. Two progenitor models, a 40 \msun\ model from \cite{WoWe95} and a 50 \msun\ model from \cite{ToUmNo07}, were considered using three different EOS: Shen \cite{ShToOy98}, LS180 \cite{LaSw91}, and LS220 \cite{LaSw91}. The simulations used the spherically symmetric relativistic treatment of gravity derived in \cite{MiSh64}. For the linear PNS analysis, the outer boundary is chosen as the PNS surface, defined as the spherically averaged $10^{11}$ g cm$^{-3}$ density contour, and the treatment of gravity is consistent with the CCSN simulation treatment of gravity, i.e., relativistic \cite{SoKuTa19}. While the treatment of gravity is relativistic, the linear PNS analysis was performed using the relativistic Cowling approximation and did not include metric perturbations.

In Sotani et al. 2021 \cite{SoTaTo21}, the 20 \msun\ progenitor from \cite{WoHe07} and the 2.9 \msun\ helium star of \cite{MoMaTo17} are evolved in two-dimensional CCSN simulations using the \small{3DnSNe-ISDA} code \cite{TaKoSu12,TaKoSu14,TaKoSu16}. The treatment of gravity was pseudo-Newtonian. The 20 \msun\ progenitor model was evolved using four different EOS: DD2 \cite{TyRoKl10}, SFHo \cite{StHeFi13}, TGLD \cite{FuToNa17}, and TGTF \cite{ToNaTa17}. The 2.9 \msun\ model used the LS220  EOS \cite{LaSw91}. The linear PNS analysis used a relativistic treatment of gravity in the relativistic Cowling approximation and applied boundary conditions at the PNS surface defined as the spherically averaged $10^{11}$ g cm$^{-3}$ density contour \cite{SoKuTa19}.

In Sotani et al. 2025 \cite{SoMuTa25}, the 12 \msun\ and 20 \msun\ progenitors from \cite{WoHe07} were evolved in two-dimensional CCSN simulations using the \small{CoCoNUT-FMT} code. The CCSN simulations treated gravity in the XCFC approximation and used the SFHo EOS \cite{StHeFi13}. Comparisons were made to simulations that used pseudo-Newtonian gravity, but no new universal relations were derived from these simulations. They performed linear PNS analyses that used similar methods as in Sotani et al. 2021 \cite{SoTaTo21}; i.e., relativistic gravity, both with and without metric perturbations. The methods for the analysis with metric perturbations is outlined in \cite{SoTa20b}. In this work we consider the new universal relations derived using the linear PNS analysis that included metric perturbations. They used the PNS surface defined as the spherically averaged $10^{11}$ g cm$^{-3}$ density contour as the outer boundary for their linear PNS analysis.

In Mori et al. 2023 \cite{MoSuTa23}, the 9.6 \msun, zero metallicity progenitor from \cite{HeWo10} is evolved for 20 seconds after bounce using the one-dimensional \small{GR1D} code \cite{OcOt10,Oconnor15}. The treatment of gravity for the \small{GR1D} code is spherically symmetric and relativistic, and the EOS used is DD2 \cite{TyRoKl10}. The linear PNS analysis used the same GREAT code as in Torres-Forné et al. 2021 \cite{ToCeMu21} with gravity treated in the CFC approximation and metric perturbations included. The outer boundary condition was set as the PNS surface defined as the spherically averaged $10^{11}$ g cm$^{-3}$ density contour.

In Rodriguez et al. 2023 \cite{RoRaCh23}, PNS models were obtained from the three-dimensional CCSN simulations conducted in \cite{RaMoBu19}, which,in turn, used seven progenitors from \cite{SuErWo16} with zero age main sequence (ZAMS) masses of 9--13 \msun, 19 \msun, and 60 \msun\ as well as the 25 \msun\ progenitor from \cite{SuWoHe18}. These progenitors were evolved using the SFHo EOS \cite{StHeFi13}, and the treatment of gravity was pseudo-Newtonian. They used the CFC approximation for their treatment of gravity in their linear PNS analysis and also used the relativistic Cowling approximation. The PNS surface defined as the spherically averaged $10^{10}$ g cm$^{-3}$ density contour was used as the outer boundary for the linear PNS analysis. 

\subsection{Applying Universal Relations to \small{CHIMERA Data}}
\label{subsec: applying UR}

To investigate the universal relations described above, we constructed spectrograms from the gravitational wave signals produced by the \textsc{Chimera} simulations. Spectrograms were constructed using the method detailed in section II.C in \cite{MuCaMe24}. For each model, we used the Kaiser windowing function with a shape factor of $\beta=6$. In the D15 and E-series models we used 45 ms window segments, with window segments overlapping adjoining segments by $94.\bar{6}\%$ for D15 and $93.\bar{3}\%$ for both E-series models. This resulted in an effective window width of 2.4 ms for D15 and 3 ms for both E-series models. The D25 model, with a shorter overall signal, used smaller, 25 ms windows with $96\%$ segment overlap for an effective window width of 1 ms. These values were chosen to keep a reasonable resolution bandwidth (RBW); i.e., the minimum frequency difference needed to resolve nearby signals, for the D-series models. We determine this quantity as the equivalent noise bandwidth, determined by the total number of windowing segments and sampling frequency (5000 Hz), divided by the total time length of the signal. For both D-series models the RBW was 25 Hz, and for the E-series models it was 7.3 Hz and 8.2 Hz for E15-LSBCK and E15-SFHo, respectively.

From each spectrogram, we determine the peak frequencies of the high frequency component of the gravitational wave signal as the frequencies in each time window that contain the maximum power spectral density. In \cite{MuMeLe25}, a linear perturbative analysis of the PNS found that the ${}^2g_2$, ${}^2g_1$, and ${}^2f$ oscillation mode frequencies of the PNS showed strong agreement with the frequencies with the greatest power spectral density in the high frequency region of the D-series spectrograms. Following the nomenclature introduced in \cite{MuBrRi26}, we call the collection of frequencies with the greatest power spectral density in the high frequency region of the spectrogram the $g$-/$f$-mode feature (gfF). We restrict the search for peak frequencies of the gfF to be above 250 Hz, and we define the start of the gfF as the time at which some frequency bin contains at least 1\% of the maximum spectral density of the spectrogram for all frequencies above 50 Hz. This is the same definition as used in \cite{MuCaMe24}.

After constructing spectrograms, we examine each universal relation through two different methods. The first method uses the evolution of the mass and radius of the PNS, as determined from \textsc{Chimera} simulation data, to compute the PNS compactness
\begin{equation}
    x_1=\frac{M_{PNS}}{R_{PNS}},
\end{equation}
PNS surface gravity
\begin{equation}
    x_2=\frac{M_{PNS}}{R^2_{PNS}}, \label{eq:surfg}
\end{equation}
and PNS mean density
\begin{equation}
    x_3=\sqrt{\frac{M_{PNS}}{R^3_{PNS}}}\label{eq:meanden}
\end{equation}
as functions of time. These quantities are computed with the PNS surface defined as either the spherically averaged $10^{11}$ g cm$^{-3}$ or $10^{10}$ g cm$^{-3}$ density contours. We insert $x_n$ into Eq. \eqref{eq:UR_base} to determine the predicted PNS oscillation mode frequencies for each universal relation. We overlay the predicted PNS oscillation frequencies onto the constructed spectrograms and compare them to the gfF peak frequencies.

Instead of solving Eq. \ref{eq:UR_base} for the PNS oscillation frequency $f$, our second analysis method solves for $x_n$ in each universal relation through a root solve of the equation
\begin{equation}
    0=c_1+c_2\mathrm{ln}(x)+c_3x+c_4x^2+c_5x^3-f,\label{eq:UR_inv},
\end{equation}
where $f$ is input as the gfF peak frequency determined as described above. In this way, we determine a predicted $x_n$ for each universal relation as a function of time. We then compare this to the PNS surface gravity and mean density evolutions computed directly from the \textsc{Chimera} simulation data by Eqs. \eqref{eq:surfg} and \eqref{eq:meanden}. In using this second analysis, it is important to note that our spectrograms represent the ideal case of a gravitational wave detection in the absence of noise. An example of determining the PNS surface gravity using the universal relation of Torres-Forné et al. 2019 \cite{ToCeOb19} in the presence of gravitational wave detector noise noise is shown in \cite{PoMu22}.

We note that our second analysis method is just one way that could be used to extract PNS properties from a CCSN gravitational wave detection. An alternative method is proposed and investigated in Bizouard et al. 2021 \cite{BiMaTo21}, where a function is derived to explicitly predict the PNS surface gravity as a function of the gfF frequency, rather than the other way around, as in Eq. \eqref{eq:UR_base}. This function was derived using a subset of the one-dimensional models considered in Torres-Forné et al. 2021 \cite{ToCeMu21}, and the PNS analysis was performed by the GREAT code \cite{ToCePa18,ToCePa19}. While the relations of Torres-Forné et al 2021 \cite{ToCeMu21} were derived using the PNS surface defined as the spherically averaged $10^{10}$ g cm$^{-3}$ density contour, in Bizouard et al. 2021 \cite{BiMaTo21} the PNS surface was defined as the spherically averaged $10^{11}$ g cm$^{-3}$ density contour. We include the function derived in Bizouard et al. 2021 \cite{BiMaTo21} in our analysis, and it is given by
\begin{equation}
    \frac{M_{PNS}}{R^2_{PNS}}=\left(2.00\times10^{-6}\right)f+\left(-1.64\times10^{-9}\right)f^2+\left(2.03\times10^{-12}\right)f^3,
\end{equation}
where $f$ is the gfF frequency in Hz and the units on each constant are such that the surface gravity has units of \msun\ km$^{-2}$. Another novel method for extracting PNS parameters is investigated in \cite{WoFrMi23}. However, that approach utilizes characteristic frequency matching to PNS parameters instead of the frequency evolution we use in this work; therefore, it is not considered here.

\section{Results}
\label{sec:results}

\begin{figure}
    \centering
    \includegraphics[width=0.8\linewidth]{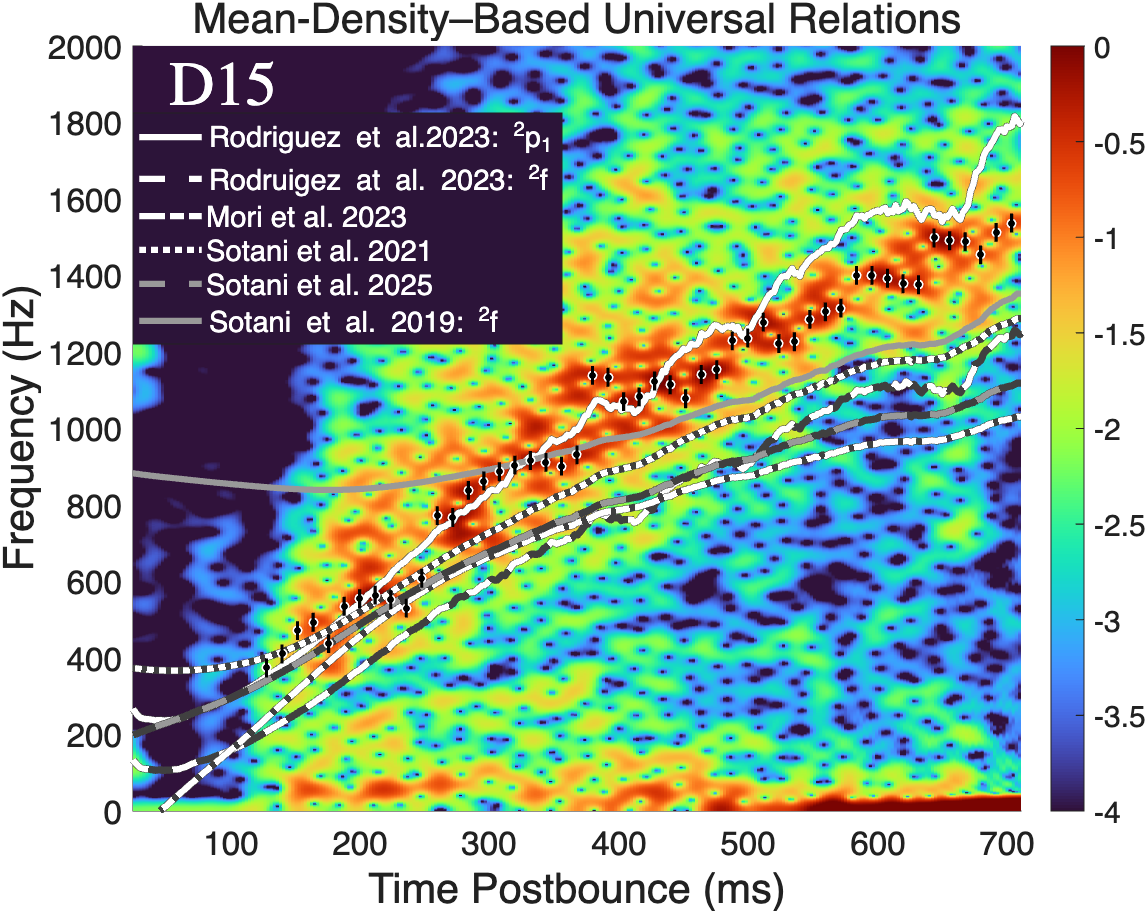}
    \caption{Spectrogram of the $h_+$ strain gravitational wave emission from model D15, with the color axis representing the logarithm of the power spectral density $\mathrm{log}_{10}(P)$. In white-edged black dots, we show the frequency with the maximum power spectral density in a time window spaced every 12 ms. A vertical black bar centered on each circle shows the range of the RBW, as defined in the text. Overlaid on each spectrogram are lines showing the universal relations based on mean density of the PNS,  \textit{i.e.}, $\sqrt{MR^{-3}}$. The solid white line, dashed white line, dash-dotted white line, dotted white line, and solid gray line correspond to the relations derived by Rodriguez et al. 2023 \cite{RoRaCh23} for ${}^2p_1$ and ${}^2f$ modes, Mori et al. 2023 \cite{MoSuTa23}, Sotani et al. 2021 \cite{SoTaTo21}, and Sotani and Sumiyoshi 2019 \cite{SoSu19}, respectively.}
    \label{fig:rho_spec_15}
\end{figure}

In Fig. \ref{fig:rho_spec_15} we show the spectrogram for D15 with the PNS mean-density--based universal relations overlaid in different line types. The peak frequencies of the gfF are shown as white-edged black dots every 12 ms from the start of the gfF. The universal relations from Sotani et al. 2021 \cite{SoTaTo21} and Sotani et al. 2025 \cite{SoMuTa25} approximately track the gfF frequencies from $\sim$130--240 ms after bounce, only slightly underestimating the frequencies. At later times, these same relations severely underestimate the gfF frequencies. The ${}^2f$ mode relation from Rodriguez et al. 2023 \cite{RoRaCh23} is never particularly close to the gfF. The relations from Mori et al. 2023 \cite{MoSuTa23} and Sotani and Sumiyoshi 2019 \cite{SoSu19} are briefly close to the gfF at $\sim$240 ms and $\sim$300--370 ms after bounce respectively, but otherwise do not track the gfF. The ${}^2p_1$ mode relation of Rodriguez et al. 2023 \cite{RoRaCh23} tracks the peak gfF frequencies generally quite well, with deviations only beginning to occur after $\sim$500 ms after bounce. It is not clear if the relation will continue to overestimate the gfF frequencies with brief periods of close approximation, as from $\sim$610--650 ms after bounce, or if the relation will continue to diverge at later times. 

\begin{figure}
    \centering
    \includegraphics[width=0.8\linewidth]{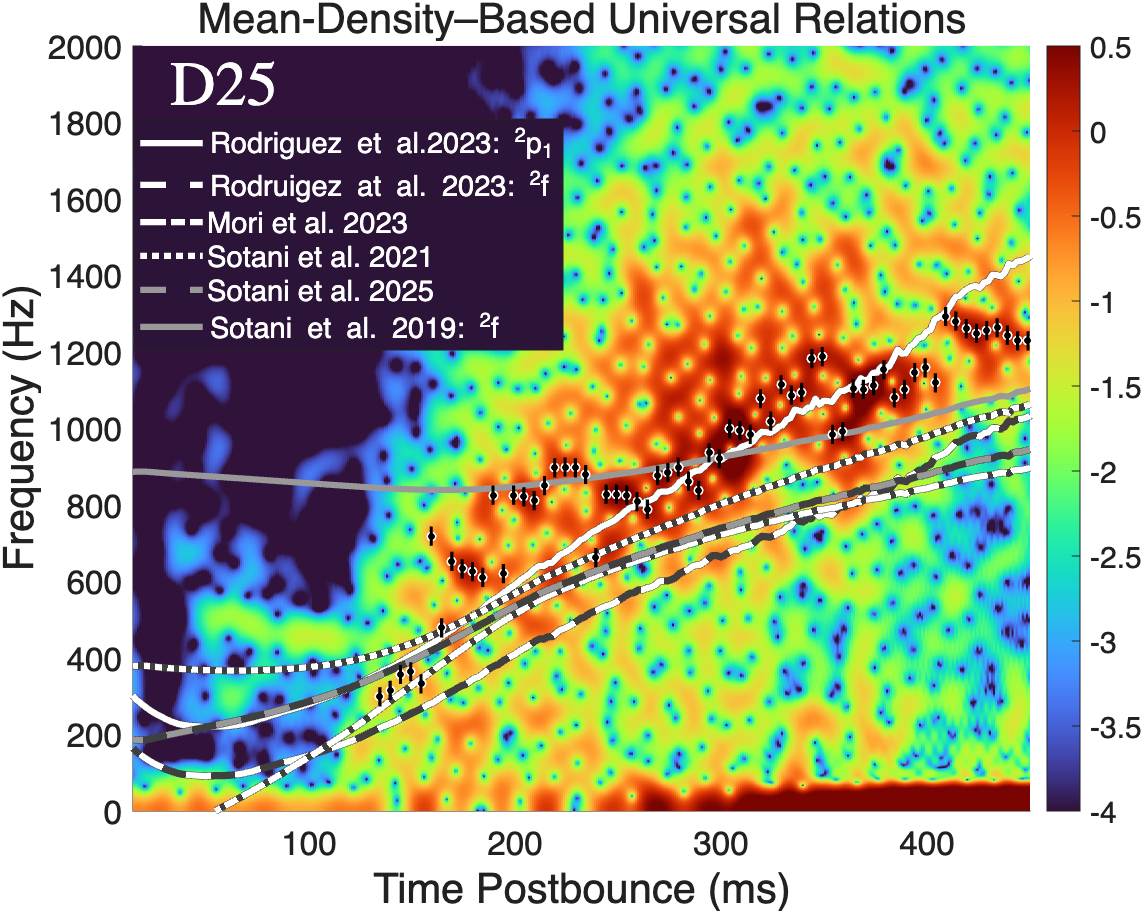}
    \caption{Spectrogram of the $h_+$ strain gravitational wave emission from model D25 with corresponding results overlaid as in Fig. \ref{fig:rho_spec_15}, with the same legend. The spacing for the gfF frequencies, in white-edged black dots, is 10 ms.}
    \label{fig:rho_spec_25}
\end{figure}

For the D25 model, Fig. \ref{fig:rho_spec_25} shows the spectrogram for the gravitational wave signal of D25 and we see more stochasticity in the gfF frequencies as shown by the white-edged black dots every 5 ms. In contrast to the D15 model, this leads to a short period where the relations of Sotani et al. 2025 \cite{SoMuTa25}, Mori et al. 2023 \cite{MoSuTa23}, and the ${}^2p_1$ mode of Rodriguez et al. 2023 \cite{RoRaCh23} approximately track the gfF from $\sim$130--150 ms after bounce. From $\sim$180--200 ms after bounce, no universal relation tracks the gfF. After this, the results become more similar to the D15 case. There is a slightly earlier period over which the relation from Sotani and Sumiyoshi 2019 \cite{SoSu19} approximately tracks the gfF, from $\sim$190--300 ms after bounce, and beginning $\sim$260 ms after bounce the ${}^2p_1$ mode of Rodriguez et al. 2023 \cite{RoRaCh23} approximately tracks the gfF. Again, it is not clear if the relation of Rodriguez et al. 2023 \cite{RoRaCh23} will continue to approximately track the gfF, or if it is beginning to diverge beginning at $\sim$400 ms after bounce.

\begin{figure}
    \centering
    \includegraphics[width=0.8\linewidth]{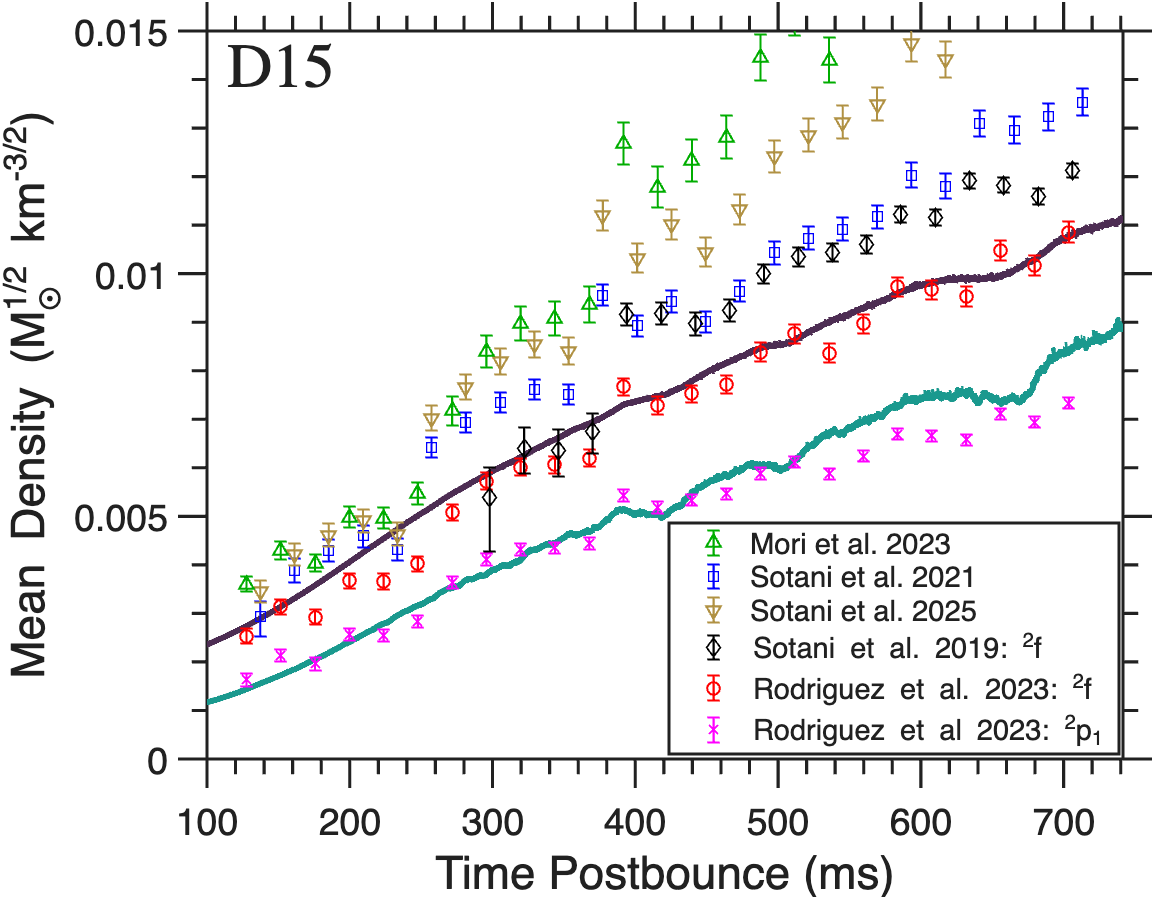}
    \caption{PNS mean density as a function of time for D15. The mean density from \textsc{Chimera} data with the PNS surface defined at the $10^{11}$ g cm$^{-3}$ density contour is shown as a dark purple line. The mean density with the PNS surface defined at the $10^{10}$ g cm$^{-3}$ density contour is shown in the lighter teal line. The mean densities predicted by each universal relation using gfF frequencies are plotted in different colored symbols every 24 ms; i.e., every other frequency shown as white-edged black dots in Fig. \ref{fig:rho_spec_15}. The green triangles, blue rectangles, black diamonds, red circles, and magenta x's correspond to the relations from Mori et al. 2023 \cite{MoSuTa23}, Sotani et al. 2021 \cite{SoTaTo21}, Sotani and Sumiyoshi 2019 \cite{SoSu19}, and Rodriguez et al. 2023 \cite{RoRaCh23} for their ${}^2f$ and ${}^2p_1$ modes, respectively. Error-bars show the spread determined by adding or subtracting the RBW value from the frequency used in each universal relation.}
    \label{fig:rho_solve_15}
\end{figure}

\begin{figure}
    \centering
    \includegraphics[width=0.8\linewidth]{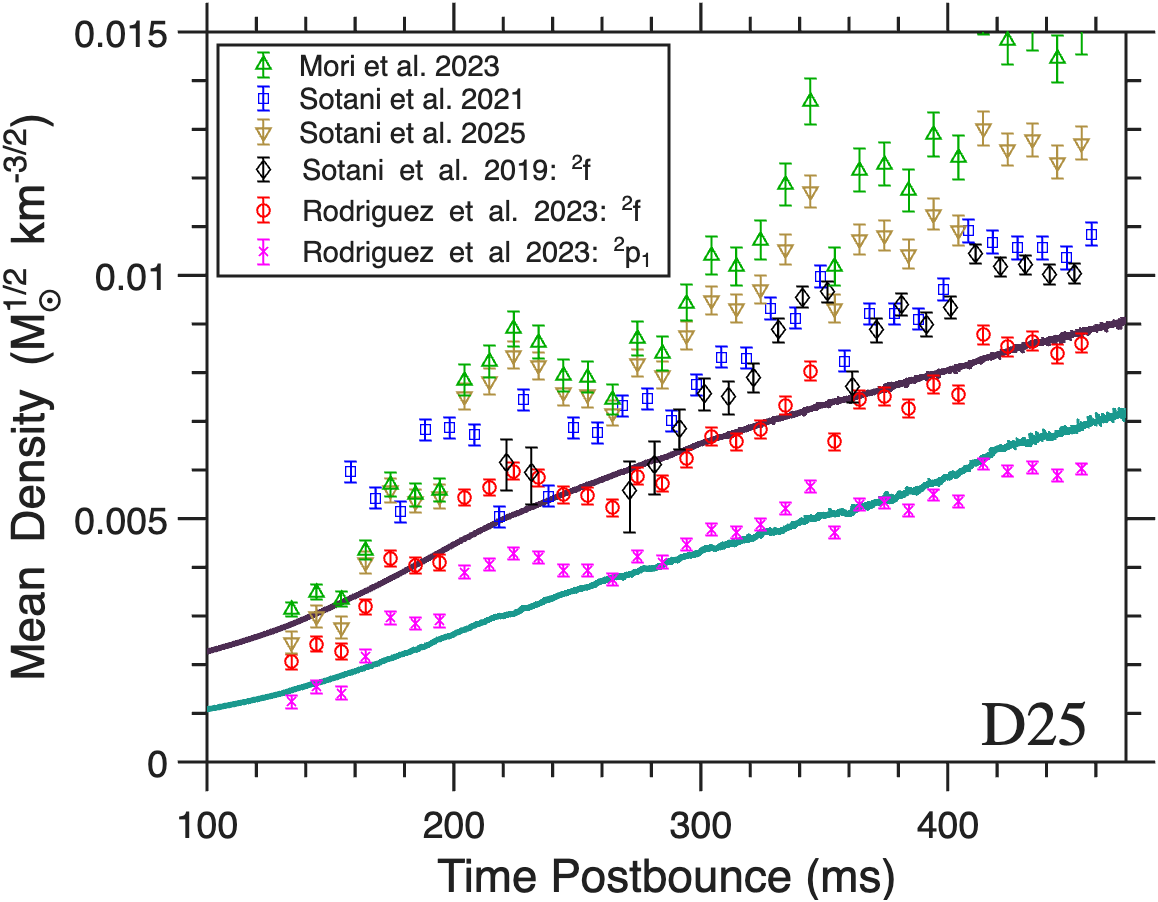}
    \caption{Same as Fig. \ref{fig:rho_solve_15} for model D25; i.e., \textsc{Chimera} data for PNS mean density with the PNS surface defined at the $10^{11}$ g cm$^{-3}$ density contour shown as a dark purple line and the mean density with the PNS surface defined at the $10^{10}$ g cm$^{-3}$ density contour shown as the lighter teal line. The predicted mean densities from each universal relation are denoted by different symbols every 10 ms with the same legend as Fig. \ref{fig:rho_solve_15}.}
    \label{fig:rho_solve_25}
\end{figure}

We show the results of our second method of analysis for both D15 and D25 in Figs. \ref{fig:rho_solve_15} and \ref{fig:rho_solve_25}. These were obtained by inputting the gfF peak frequencies determined from each spectrogram (e.g., white-edged dots in Figs. \ref{fig:rho_spec_15} and \ref{fig:rho_spec_25}) into Eq. \eqref{eq:UR_inv} and applying a root-solving algorithm to predict the mean density of the PNS. The error-bars shown in both figures represent the mean density computed using the gfF frequency plus or minus the RBW frequency. As should be expected from their predicted gfF frequencies in D15, beyond $\sim$240 ms after bounce the relations of Sotani et al. 2021 \cite{SoTaTo21}, Mori et al. 2023 \cite{MoSuTa23}, and Sotani et al. 2025 \cite{SoMuTa25} overestimate the PNS mean density. Likewise, the ${}^2f$ mode relation of Sotani and Sumiyoshi 2019 \cite{SoSu19} predicts the PNS mean density, within the error determined from the RBW of the spectrogram, from $\sim$300--370 ms and overestimates the mean density at later times. The ${}^2p_1$ mode relation of Rodriguez et al. 2023 \cite{RoRaCh23} tracks well the PNS mean density with the surface defined as the spherically averaged $10^{10}$ g cm$^{-3}$ density contour (teal line) until $\sim$500 ms after bounce, and then begins to slightly underestimate the mean density. Surprisingly, the ${}^2f$ mode relation of Rodriguez et al. 2023 \cite{RoRaCh23} approximately tracks the PNS mean density with the PNS surface defined as the spherically averaged $10^{11}$ g cm$^{-3}$ density contour after $\sim$300 ms after bounce. For the ${}^2f$ mode relation of Rodriguez et al. 2023 \cite{RoRaCh23} we do not see the same divergence from the real mean density as is seen in their ${}^2p_1$ relation. However, while this relation does track the PNS mean density, it is not the mean density it  was derived to track. It is not immediately clear if relations derived using one definition of the PNS surface can consistently use a different definition of the PNS surface to predict PNS oscillation frequencies. The results for D25, shown in Fig. \ref{fig:rho_solve_25}, are similar, with the specific times at which a universal relation tracks the mean density aligning well with the times noted for Fig. \ref{fig:rho_spec_25}. We see the same behavior with the ${}^2p_1$ mode and ${}^2f$ mode relations of Rodriguez et al. 2023 \cite{RoRaCh23} approximately tracking the mean density of the PNS, but with the later using a different definition of the PNS surface. 

\begin{figure}
    \centering
    \includegraphics[width=0.8\linewidth]{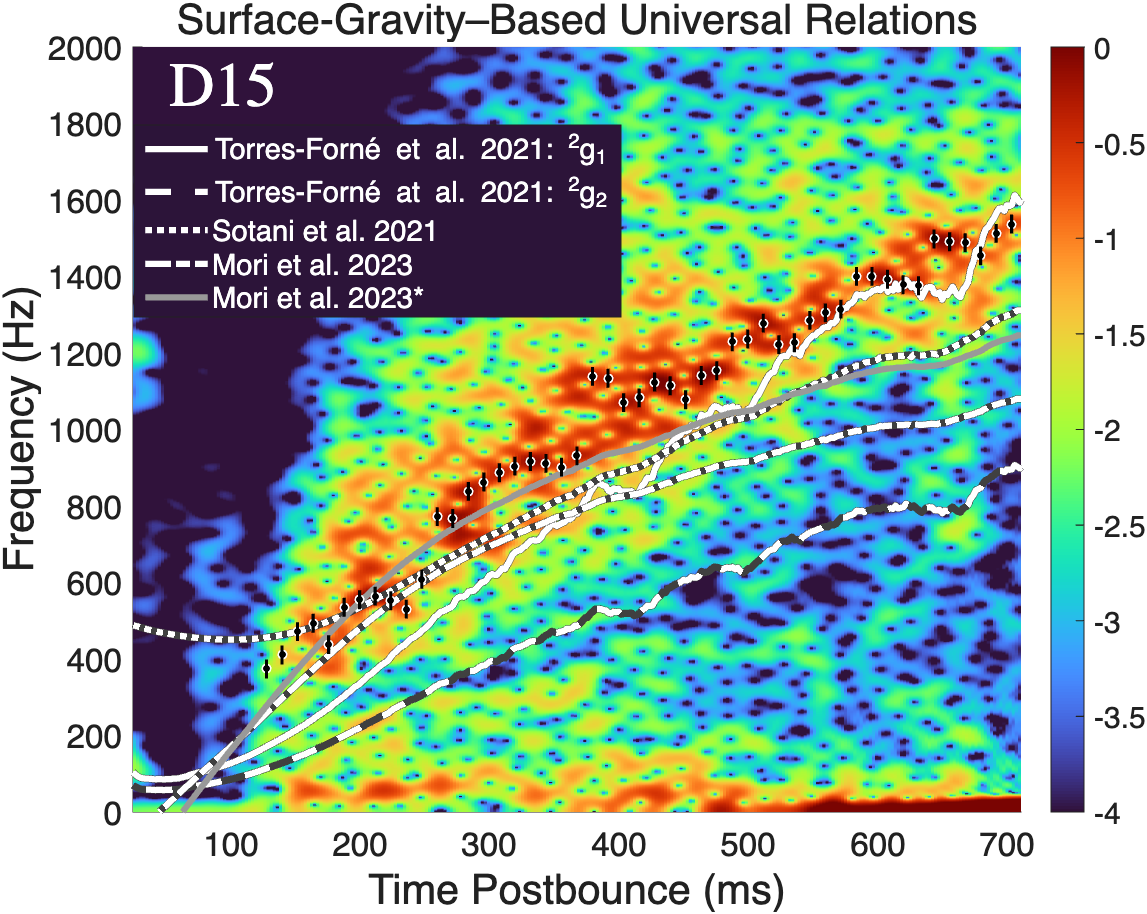}
    \caption{Spectrogram of the $h_+$ strain gravitational wave emission from D15, with the same properties as Fig. \ref{fig:rho_spec_15}, with different lines overlaid corresponding to the universal relations based on surface gravity of the PNS; \textit{i.e.}, $MR^{-2}$. The solid white line, dashed white line, dash-dotted white line, and dotted white line correspond to the relations derived by Torres-Forné et al. 2021  \cite{ToCeMu21} for their ${}^2g_1$ and ${}^2g_2$ modes, Sotani et al. 2021 \cite{SoTaTo21}, and Mori et al. 2023 \cite{MoSuTa23}, respectively. The solid gray line corresponds to the relation based on compactness; \textit{i.e.}, $MR^{-1}$, derived by Mori et al. 2023 \cite{MoSuTa23}.}
    \label{fig:sg_spec_15}
\end{figure}

\begin{figure}
    \centering
    \includegraphics[width=0.8\linewidth]{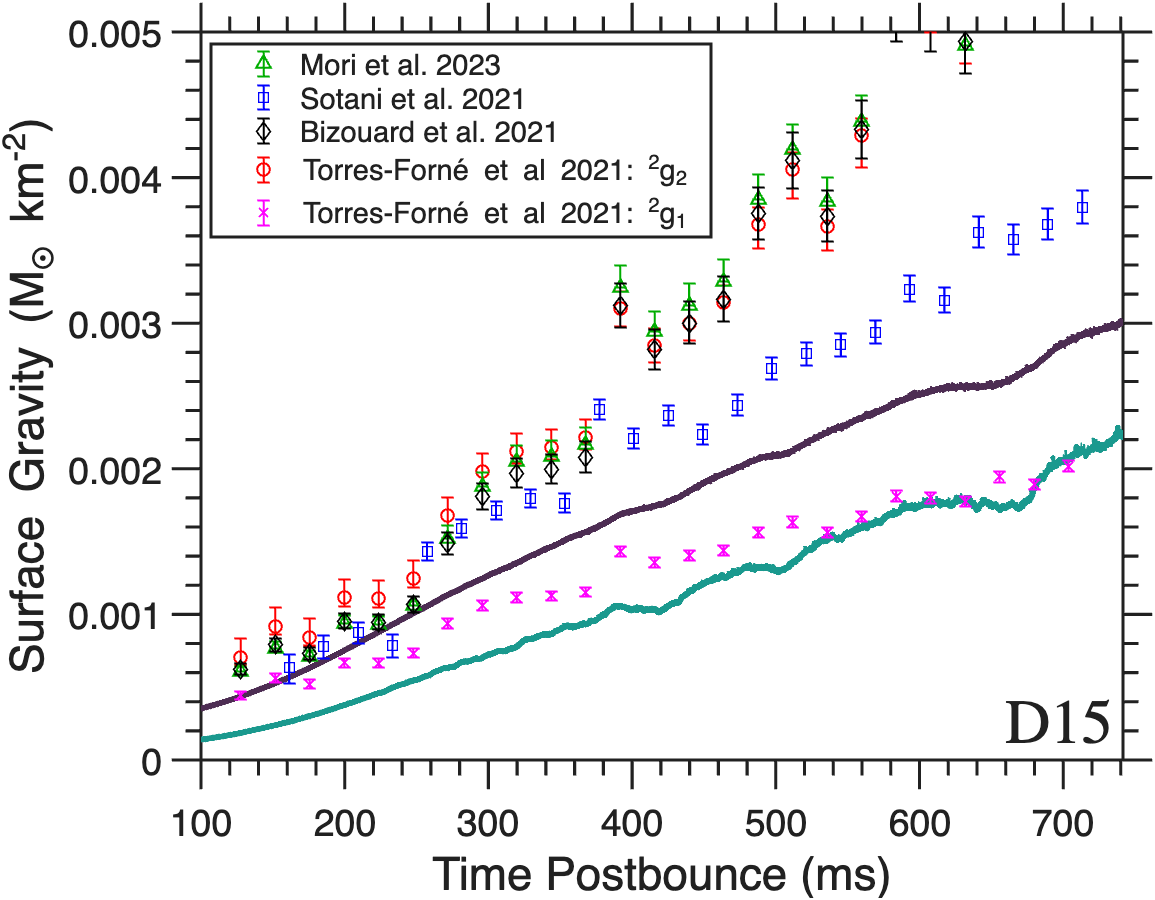}
    \caption{PNS surface gravity from \textsc{Chimera} data as a function of time for D15. The surface gravity with the PNS surface defined at the $10^{11}$ g cm$^{-3}$ density contour is shown as a dark purple line while the surface gravity with the PNS surface defined at the $10^{10}$ g cm$^{-3}$ density contour is shown in the lighter teal line. The surface gravities predicted by each universal relation using the gfF frequencies are denoted by different colored symbols every 24 ms. The green triangles, blue rectangles, black diamonds, red circles, and magenta x's correspond to the relations from Mori et al. 2023 \cite{MoSuTa23}, Sotani et al. 2021 \cite{SoTaTo21}, Bizouard et al. 2021 \cite{BiMaTo21}, and Torres-Forné et al. 2021 \cite{ToCeMu21} for their ${}^2g_2$ and ${}^2g_1$ modes, respectively. Error-bars show the spread determined by adding or subtracting the RBW value from the frequency used in each universal relation.}
    \label{fig:sg_solve_15}
\end{figure}

\begin{figure}
    \centering
    \includegraphics[width=0.8\linewidth]{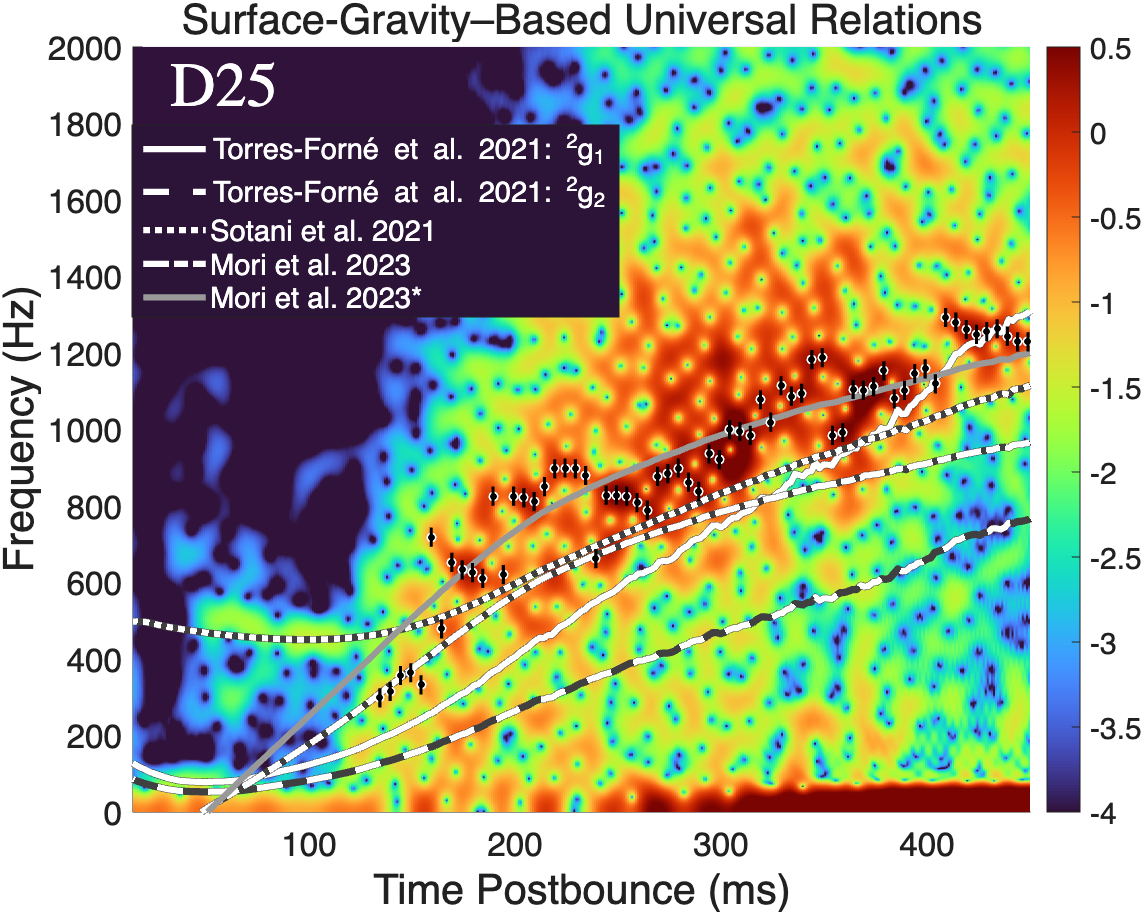}
    \caption{Spectrogram of the $h_+$ strain gravitational wave emission from D25, with the same properties as Fig. \ref{fig:rho_spec_25}, and corresponding results overlaid in lines with the same legend as in Fig. \ref{fig:sg_spec_15}.}
    \label{fig:sg_spec_25}
\end{figure}

\begin{figure}
    \centering
    \includegraphics[width=0.8\linewidth]{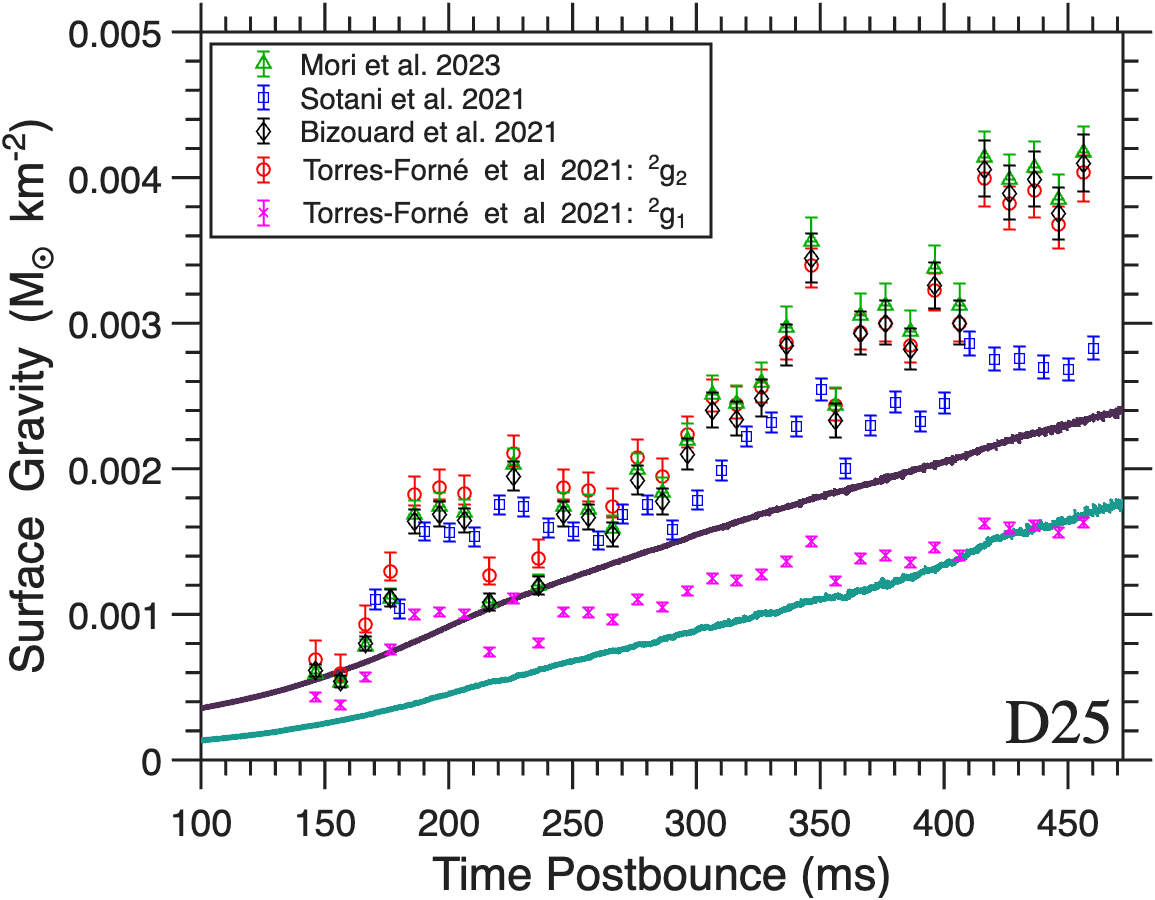}
    \caption{PNS surface gravity from \textsc{Chimera} data as a function of time for D25, with the same legend as Fig. \ref{fig:sg_solve_15}. The surface gravity with the PNS surface defined at the $10^{11}$ g cm$^{-3}$ density contour is shown as a dark purple line while the surface gravity with the PNS surface defined at the $10^{10}$ g cm$^{-3}$ density contour is shown in the lighter teal line. The surface gravities predicted by each universal relation using the gfF frequencies are denoted by different colored symbols every 10 ms.}
    \label{fig:sg_solve_25}
\end{figure}

In Figs. \ref{fig:sg_spec_15}--\ref{fig:sg_solve_25} we show the same results for those universal relations based on surface gravity, as well as the relation based on compactness from Mori et al. 2023 \cite{MoSuTa23}. In Fig. \ref{fig:sg_spec_15} for the D15 model, we see the relation from Sotani et al. 2021 \cite{SoTaTo21} approximating the gfF frequencies and, in turn in Fig. \ref{fig:sg_solve_15}, the PNS surface gravity, from $\sim$150--250 ms after bounce. During those same times, the function derived by Bizouard et al. 2021 \cite{BiMaTo21} also roughly tracks the PNS surface gravity. From $\sim$220--240 ms after bounce the surface-gravity--based relation of Mori et al. 2023 \cite{MoSuTa23} also approximates the gfF frequencies and PNS surface gravities. At later times, none of those relations track the gfF or PNS surface gravity. Beginning at $\sim$520 ms after bounce, the ${}^2g_1$ mode relation of Torres-Forné et al. 2021 \cite{ToCeMu21} approximately tracks the gfF and similarly tracks the PNS surface gravity consistent with their PNS analysis---i.e., with the PNS surface defined as the spherically averaged $10^{10}$ g cm$^{-3}$ density contour (teal line). For D25, in Fig. \ref{fig:sg_spec_25} the surface-gravity--based relation of Mori et al. 2023 \cite{MoSuTa23} matches the gfF and predicts the PNS surface gravity approximately at $\sim$140--160 ms and $\sim$210--230 ms in Fig. \ref{fig:sg_solve_25}, at which times the relation of Bizouard et al. 2021 \cite{BiMaTo21} also predicts the PNS surface gravity approximately. No other surface-gravity--based relations seem to track the gfF, or predict the PNS surface gravity, accurately until the ${}^2g_1$ mode relation of Torres-Forné et al. 2021 \cite{ToCeMu21} begins to track both at $\sim$360 ms after bounce. However, in both models the compactness $\left(M_{PNS}\ R_{PNS}^{-1}\right)$ based relation of Mori et al. 2023 \cite{MoSuTa23} is relatively close to the identified gfF frequencies until $\sim$380 ms for the D15 model and until the end of the simulation of the D25 model. 

\begin{figure}
\centering
    \includegraphics[width=0.8\linewidth]{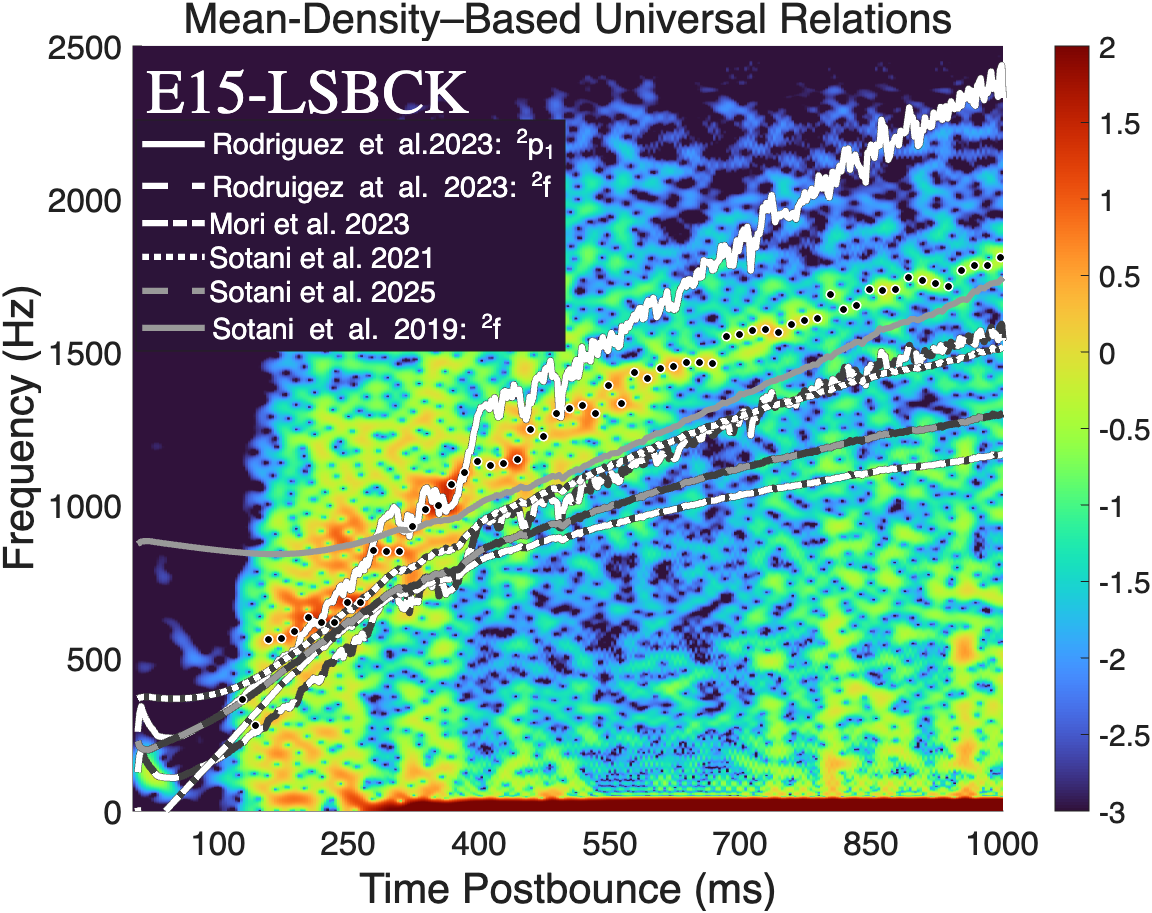}
    \caption{Spectrogram of the $h_+$ strain gravitational wave emission from model E15-LSBCK with corresponding results for mean-density--based universal relations overlaid as in Fig. \ref{fig:rho_spec_15}, with the same legend. The spacing for the gfF frequencies, in white-edged black dots, is 15 ms.}
    \label{fig:E-rho_L}
\end{figure}

\begin{figure}
\centering
    \includegraphics[width=0.8\linewidth]{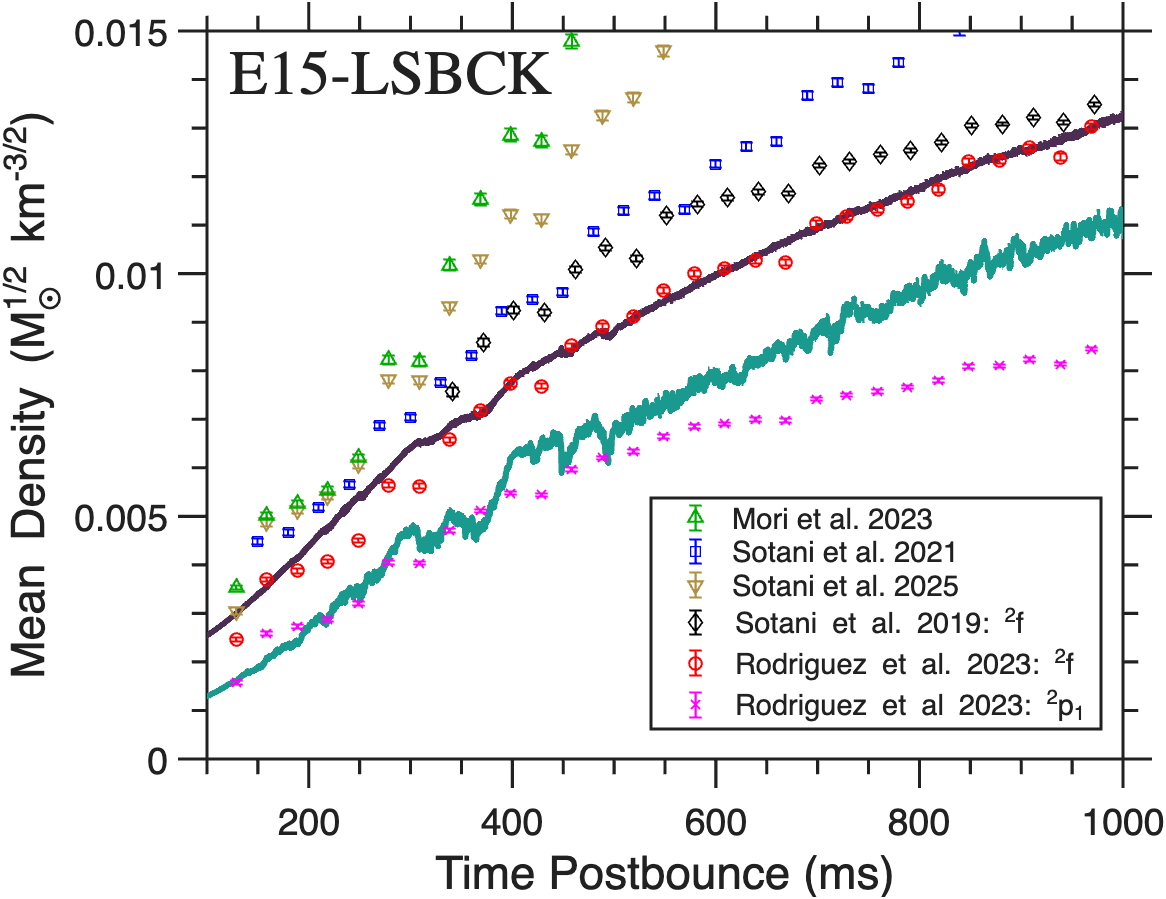}
    \caption{PNS mean density from \textsc{Chimera} data as a function of time for E15-LSBCK, with the same legend as Fig. \ref{fig:rho_solve_15}. The surface gravity with the PNS surface defined at the $10^{11}$ g cm$^{-3}$ density contour is shown as a dark purple line while the surface gravity with the PNS surface defined at the $10^{10}$ g cm$^{-3}$ density contour is shown in the lighter teal line. The mean densities predicted by each universal relation using the gfF frequencies are denoted by different colored symbols every 30 ms.}
    \label{fig:E-rho_solve_L}
\end{figure}

\begin{figure}
\centering
    \includegraphics[width=0.8\linewidth]{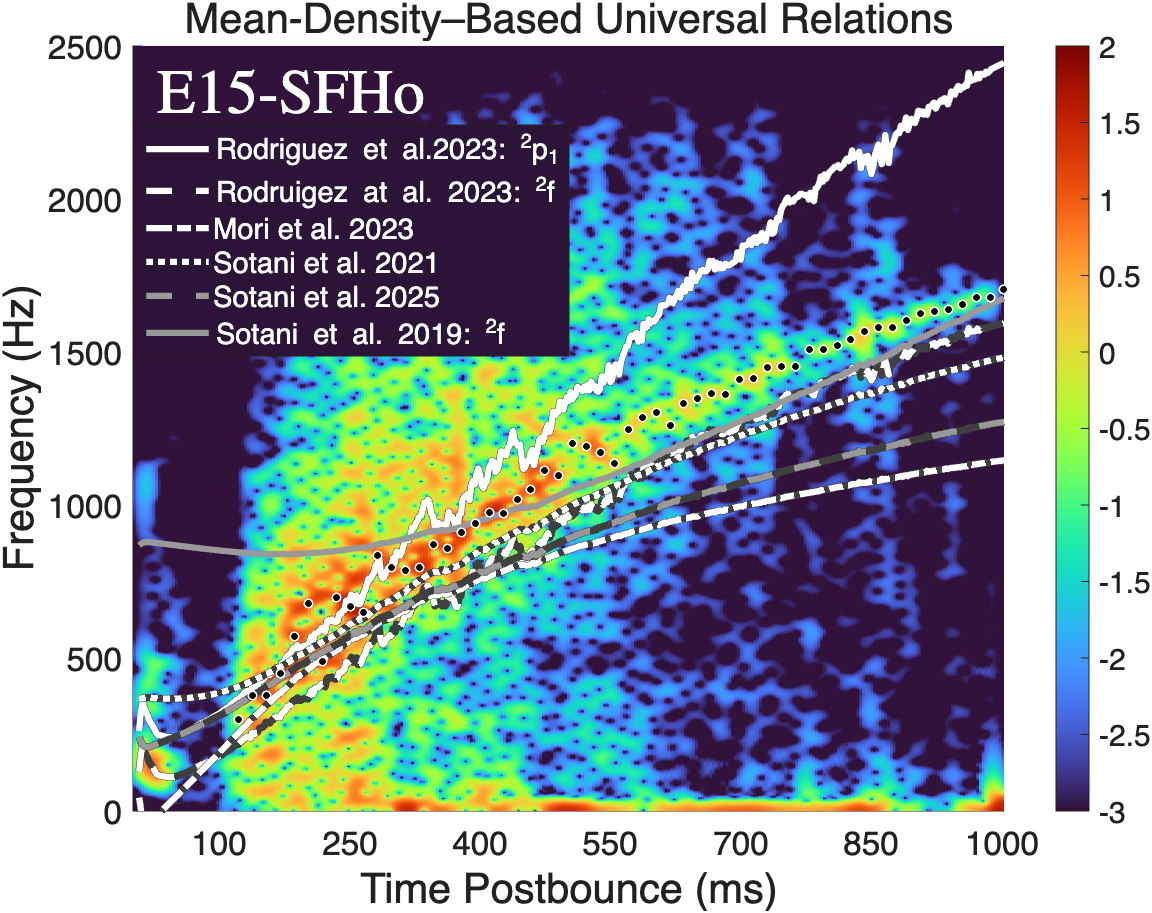}
    \caption{Spectrogram of the $h_+$ strain gravitational wave emission from model E15-SFHo with corresponding results for mean-density--based universal relations overlaid as in Fig. \ref{fig:rho_spec_15}, with the same legend. The spacing for the gfF frequencies, in white-edged black dots, is 15 ms.}
    \label{fig:E-rho_S}
\end{figure}

\begin{figure}
\centering
    \includegraphics[width=0.8\linewidth]{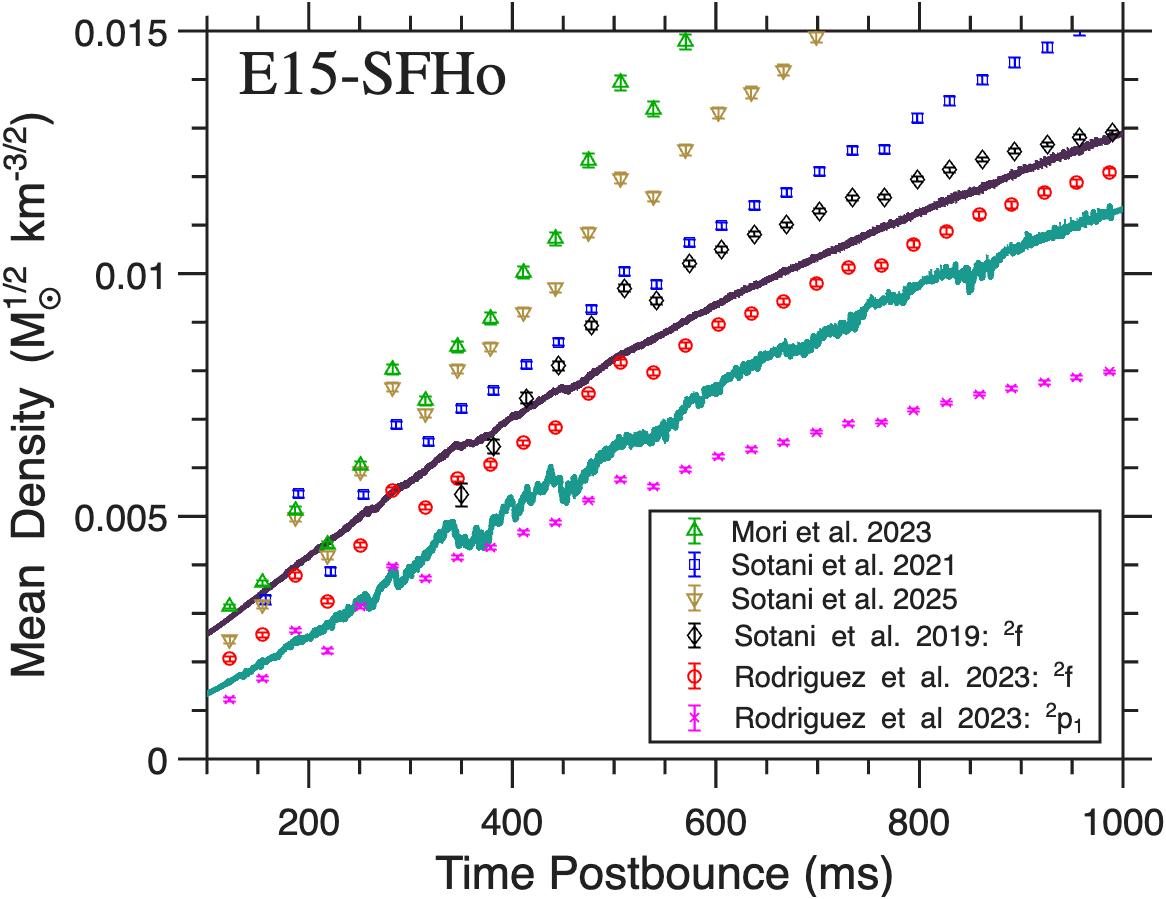}
    \caption{PNS mean density from \textsc{Chimera} data as a function of time for E15-SFHo, with the same legend as Fig. \ref{fig:rho_solve_15}. The mean density with the PNS surface defined at the $10^{11}$ g cm$^{-3}$ density contour is shown as a dark purple line, and the mean density with the PNS surface defined at the $10^{10}$ g cm$^{-3}$ density contour is shown in the lighter teal line. The mean densities predicted by each universal relation using the gfF frequencies are denoted by different colored symbols every 30 ms.}
    \label{fig:E-rho_solve_S}
\end{figure}

As only Rodriguez et al. 2023 \cite{RoRaCh23} derived their universal relations using three-dimensional CCSN simulation data, we investigate the effects of dimensionality by applying the universal relations to two, two-dimensional models from the \textsc{Chimera} E-series. These models were evolved for much longer than the D-series, but we restrict our investigation to the data produced only up to one second after bounce. In Figs. \ref{fig:E-rho_L}--\ref{fig:E-rho_solve_S} we show the results for the universal relations based on the mean density of the PNS applied to the E-series. For those relations that were derived using the PNS surface as the spherically averaged $10^{11}$ g cm$^{-3}$ density contour (i.e., Sotani and Sumiyoshi \cite{SoSu19}, Sotani et al 2021 \cite{SoTaTo21}, Sotani et al. 2025 \cite{SoMuTa25}, and Mori et al. 2023 \cite{MoSuTa23}) the results for E15-LSBCK in Figs. \ref{fig:E-rho_L} and \ref{fig:E-rho_solve_L} are qualitatively similar to the D15 model for the first $\sim$700 ms after bounce. This is true for the results in model E15-SFHo, as well, in Figs. \ref{fig:E-rho_S} and \ref{fig:E-rho_solve_S}. There is a short, initial period where the predicted frequencies are relatively close to the gfF and the predicted PNS mean densities are relatively close to the real PNS mean densities, from $\sim$130--250 after bounce for \cite{SoSu19,SoTaTo21,SoMuTa25,MoSuTa23}. Beyond $\sim$250 ms after bounce these relations diverge from the gfF and real mean density of the PNS. The relation from Sotani and Sumiyoshi 2019 \cite{SoSu19} passes through the gfF and shows approximate agreement at $\sim$300 ms after bounce for E15-LSBCK in Fig. \ref{fig:E-rho_L} and from $\sim$350--450 after bounce for E15-SFHo in Fig. \ref{fig:E-rho_S}. Conversely, at late times for both models the relation of Sotani et al. 2021 \cite{SoTaTo21} seems to have stopped diverging from the gfF and the relation of Sotani and Sumiyoshi 2019 \cite{SoSu19} has begun to approach the gfF.

For the ${}^2p_1$ mode relation of Rodriguez et al. 2023 \cite{RoRaCh23}, the predicted gfF frequencies are relatively close to the true gfF frequencies for the first $\sim$550 ms after bounce in both models in Figs. \ref{fig:E-rho_L} and \ref{fig:E-rho_S}. However, for these two-dimensional models the PNS surface at the spherically averaged $10^{10}$ g cm$^{-3}$ density contour is more volatile as reflected by the jagged features that denote rapid changes in mean density (teal line) in Figs. \ref{fig:E-rho_solve_L} and \ref{fig:E-rho_solve_S}. This makes the predicted frequencies less smooth than the identified gfF frequencies. For the data at times later than $\sim$500 ms after bounce, the predicted frequencies and PNS mean densities diverge from the gfF and true PNS mean densities significantly. This begins to happen $\sim$200 ms sooner in both E-series models as compared to the D15 model, where this divergence may just be beginning close to $\sim$700 ms after bounce. As in D15, the ${}^2f$ mode of Rodriguez et al. 2023 \cite{RoRaCh23} tracks well the mean density contained within the surface defined as the spherically averaged $10^{11}$ g cm$^{-3}$ density contour for the E15-LSBCK model. This relation tracks the mean density in E15-SFHo less well, but neither model shows signs of this relation diverging from this definition of the mean density at late times. 

\begin{figure}
\centering
    \includegraphics[width=0.8\linewidth]{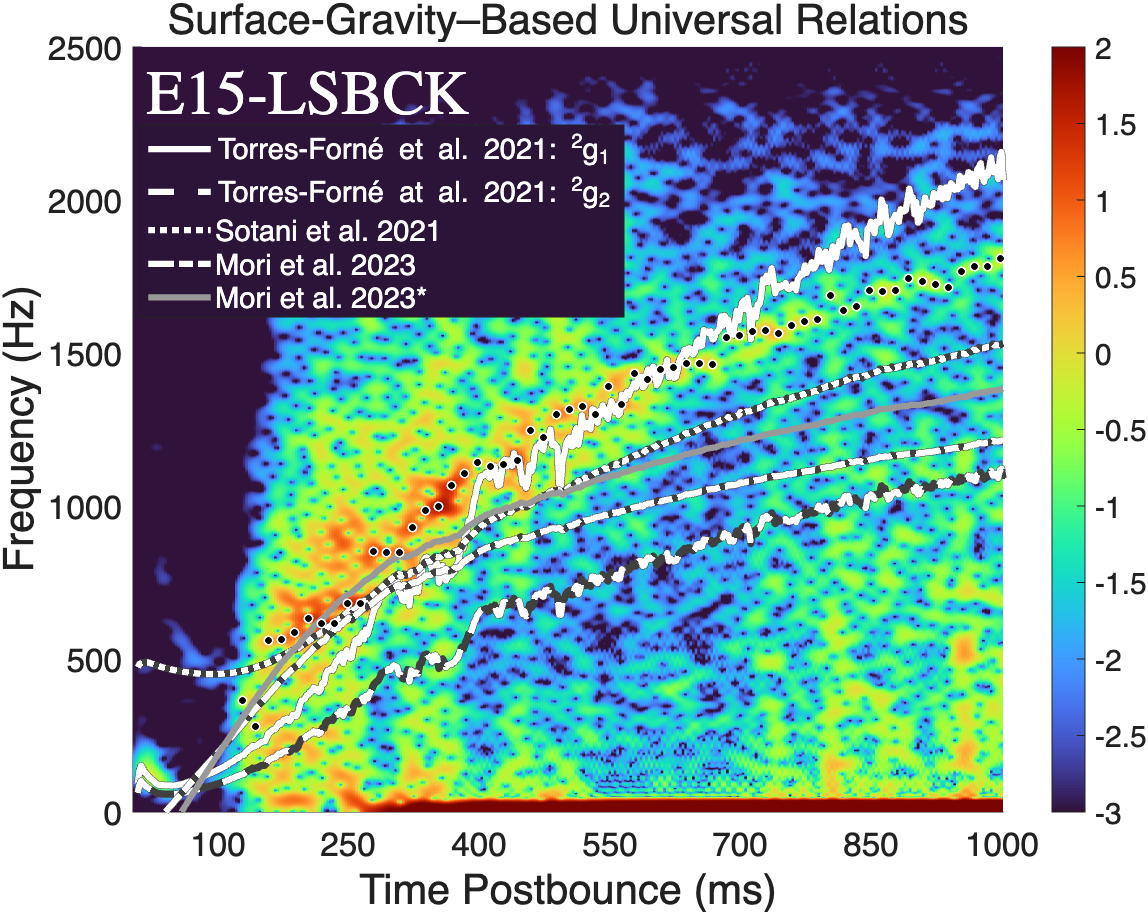}
    \caption{Spectrogram of the $h_+$ strain gravitational wave emission from E15-LSBCK, with the same properties as Fig. \ref{fig:E-rho_L}, with corresponding results for surface gravity and compactness based universal relations overlaid in lines with the same legend as in Fig. \ref{fig:sg_spec_15}.}
    \label{fig:E-sg_L}
\end{figure}

\begin{figure}
\centering
    \includegraphics[width=0.8\linewidth]{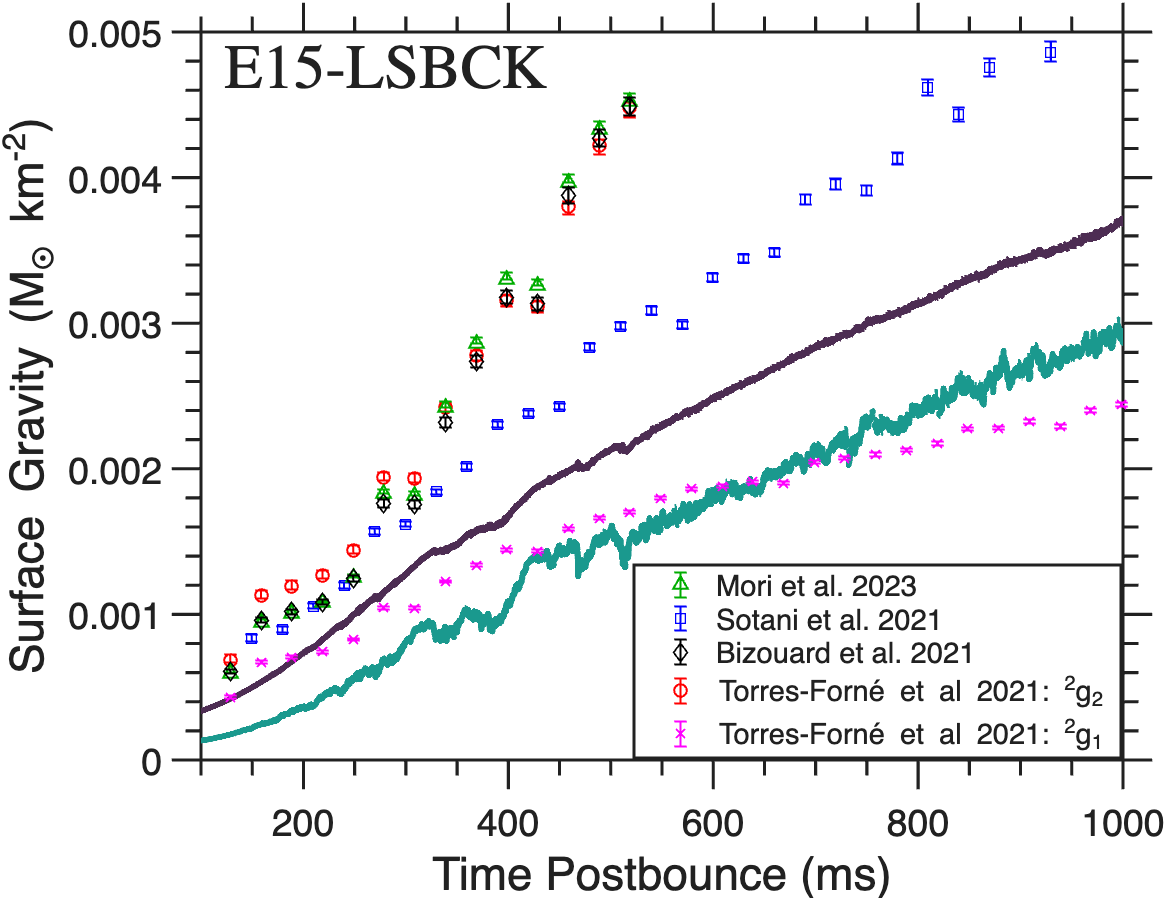}
    \caption{PNS surface gravity from \textsc{Chimera} data as a function of time for E15-LSBCK, with the same legend as Fig. \ref{fig:sg_solve_15}. The surface gravity with the PNS surface defined at the $10^{11}$ g cm$^{-3}$ density contour is shown as a dark purple line while the surface gravity with the PNS surface defined at the $10^{10}$ g cm$^{-3}$ density contour is shown in the lighter teal line. The surface gravities predicted by each universal relation using the gfF frequencies are denoted by different colored symbols every 30 ms.}
    \label{fig:E-sg_solve_L}
\end{figure}

\begin{figure}
\centering
    \includegraphics[width=0.8\linewidth]{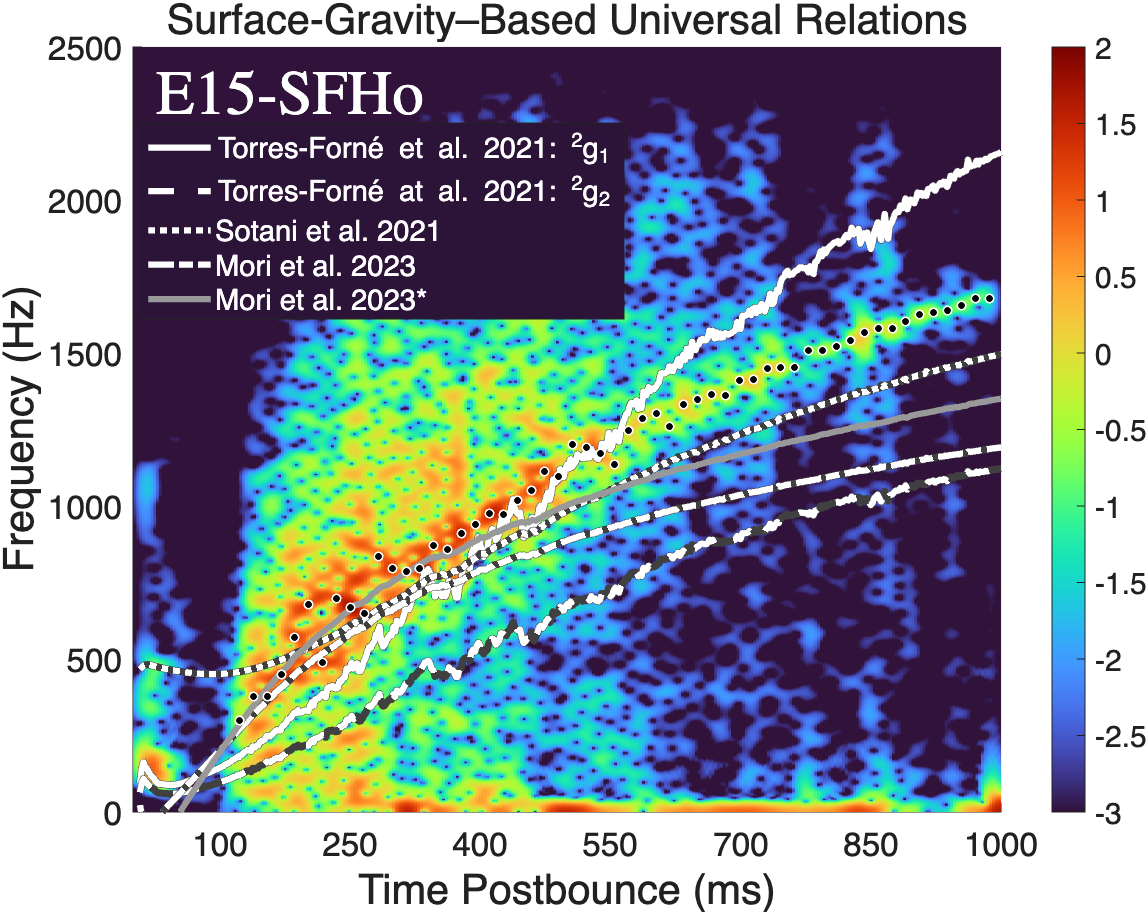}
    \caption{Spectrogram of the $h_+$ strain gravitational wave emission from E15-SFHo, with the same properties as Fig. \ref{fig:E-rho_S}, with corresponding results for surface gravity and compactness based universal relations overlaid in lines with the same legend as in Fig. \ref{fig:sg_spec_15}.}
    \label{fig:E-sg_S}
\end{figure}

\begin{figure}
\centering
    \includegraphics[width=0.8\linewidth]{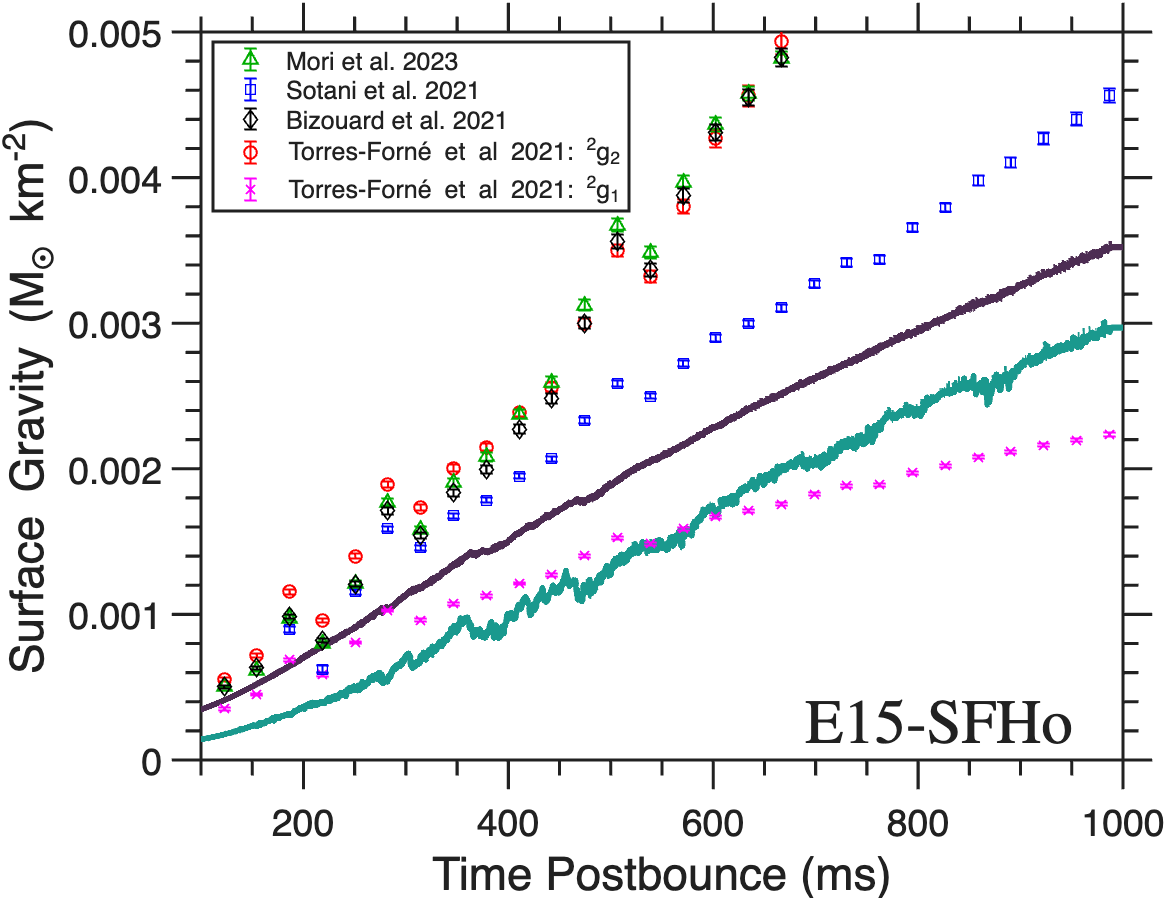}
    \caption{PNS surface gravity from \textsc{Chimera} data as a function of time for E15-SFHo, with the same legend as Fig. \ref{fig:sg_solve_15}. The surface gravity with the PNS surface defined at the $10^{11}$ g cm$^{-3}$ density contour is shown as a dark purple line while the surface gravity with the PNS surface defined at the $10^{10}$ g cm$^{-3}$ density contour is shown in the lighter teal line. The surface gravities predicted by each universal relation using the gfF frequencies are denoted by different colored symbols every 30 ms.}
    \label{fig:E-sg_solve_S}
\end{figure}

In Figs. \ref{fig:E-sg_L}--\ref{fig:E-sg_solve_S} we present the results for surface gravity and compactness based relations for the E-series. As with the mean-density--based relations, we largely see the same qualitative results between the E-series models and the D15 model. From $\sim$130--250 ms after bounce Fig. \ref{fig:E-sg_L} shows the relations of Sotani et al. 2021 \cite{SoTaTo21} and Mori et al. 2023 \cite{MoSuTa23} approximately tracking the evolution of the gfF, with the relations being closer to the identified gfF frequencies for E15-SFHo across those same times in Fig. \ref{fig:E-sg_S}. The compactness based relation of Mori et al. 2023 \cite{MoSuTa23} begins to diverge earlier in the E-series models than in D15, beginning at $\sim$300 ms after bounce for E15-LSBCK in Figs \ref{fig:E-sg_L} and $\sim$360 ms after bounce for E15-SFHo in \ref{fig:E-sg_S}. After a short period of approximately tracking the gfF and predicting the true PNS surface gravity, in Figs. \ref{fig:E-sg_solve_L} and \ref{fig:E-sg_solve_S}, from $\sim$400--600 ms after bounce, the ${}^2g_1$ mode relation of Torres-Forné et al. 2021 \cite{ToCeMu21} begins to diverge. It is unclear if the same divergence will occur at later times in the D15 model, but it has not started to diverge by $\sim$700 after bounce in Fig. \ref{fig:sg_spec_15}.

\section{Conclusions}
\label{sec:conclusion}

This work shows the application of various relations that connect PNS properties to the gravitational wave signal generated from CCSN simulations. The ultimate goal of these relations is to present a connection that is universal across progenitor properties, \textit{e.g.}, mass, rotation, metallicity, magnetic field strength and topology, etc. In addition to being universal across progenitors, the ideal relation would be universal across simulations as well. To some extent, a universality across simulations should not be expected due to differences in the physics included and numerical methods implemented. However, for simulations with similar sophistication, both physically and numerically, a universal relation that fits similarly across simulations could be expected. When applied to our simulations, with the exception of the ${}^2p_1$ mode relation from Rodriguez et al. 2023 \cite{RoRaCh23}, these universal relations do not seem to fit our data well.

Some disparities between the CCSN simulation gravitational wave frequencies and the universal relation frequencies should be expected. As shown by both \cite{West20} and \cite{SoMuTa24}, differing treatments of gravity across PNS modal analyses and CCSN simulation will result in gravitational wave frequencies that do not match. Additionally, for PNS modal analyses with relativistic treatments of gravity there is a difference in the predicted gravitational wave frequencies from analyses that use the relativistic Cowling approximation (i.e., no metric perturbations) and those that do not \cite{SoMuTa25}. In the case of CCSN simulations using a pseudo-Newtonian, effective relativistic gravitational potential (e.g, \textsc{Chimera} simulations), a PNS modal analysis using a relativistic treatment of gravity will predict gravitational wave frequencies lower than those computed from the simulation, by as much as 15\% \cite{WeOcSu19}. All relations considered in this paper, indeed, all universal relations the authors are aware of, were derived from modal analyses that utilized some relativistic treatment of gravity. 

For D15, some relations do produce frequencies close to our peak gfF frequencies for a short time after the gfF develops, from $\sim$100--300 ms after bounce: the mean density and surface gravity relations of Sotani et al. 2021 \cite{SoTaTo21}, as well as the surface gravity and compactness relations from Mori et al. 2023 \cite{MoSuTa23}. At late times, $\sim$ 500--700 ms after bounce, the surface-gravity--based relationship of Torres-Forné et al. 2021 \cite{ToCeMu21} for their ${}^2g_1$ mode is also close to the simulation values of the gfF. The mean density relationship of Rodriguez et al. 2023 \cite{RoRaCh23} for their ${}^2p_1$ mode tracks the gfF for almost the entire signal, only deviating significantly above the gfF between $\sim$500--625 ms after bounce. Out of all the relations considered in this work, the relations of Rodriguez et al. 2023 \cite{RoRaCh23} are based on CCSN simulation data most similar to the \textsc{Chimera} D-series simulations. Importantly, both sets of simulations are fully three dimensional and use a similar pseudo-Newtonian treatment of gravity, in addition to having similar levels of physical sophistication. Nonetheless, the PNS modal analysis used to derive the universal relations of Rodriguez et al. 2023 \cite{RoRaCh23} used a relativistic treatment of gravity. This mismatch in gravity treatments may contribute to why it is the ${}^2p_1$ mode universal relation of Rodriguez et al. 2023 \cite{RoRaCh23} that tracks the gfF in D15 instead of any of modes that were found to track the gfF (i.e., ${}^2g_2$, ${}^2g_1$, and ${}^2f$) in \cite{MuMeLe25}, where a consistent treatment of pseudo-Newtonian gravity across CCSN simulation and PNS modal analysis was used. Accounting for the different classification schemes between these studies---i.e., the ${}^2p_1$ (${}^2f$) mode in Rodriguez et al. 2023 \cite{RoRaCh23} is made up of the ${}^2f$ (${}^2g_1$) mode at early times and the ${}^2p_1$ (${}^2f$) mode at late times in the generalized Cowling nomenclature classification scheme used in \cite{MuMeLe25}---we would expect the ${}^2f$ mode universal relation from Rodriguez et al. 2023 \cite{RoRaCh23} to align well with the gfF. Interestingly, we do see that using this ${}^2f$ mode universal relation to solve for the PNS mean density given the gfF peak frequencies tracks the PNS mean density with the PNS surface defined at the mean $10^{11}$ g cm$^{-3}$ contour beginning $\sim$280 ms after bounce in Fig. \ref{fig:rho_solve_15}. However, this is not the mean density this universal relation intended to track as it was derived from a PNS modal analysis that set the PNS surface boundary to be the mean $10^{10}$ g cm$^{-3}$ contour. We thus have a mode that does not match the in-depth modal analysis from \cite{MuMeLe25} that tracks the gfF well, and a mode that agrees with \cite{MuMeLe25} that does track the PNS mean density well but not the PNS mean density it was derived to track. Does this make the universal relation successful, or not?

For the D25 model, the universal relations do not show the initial, moderate agreement with the simulation gfF. The more significant accretion on the PNS that occurs in this model leads to more stochastic gravitational wave emission \cite{MeMaLa23}, causing more stochastic peak gfF frequencies as well. In this model, explosion begins in earnest $\sim$250 ms after bounce, at which point accretion onto the PNS is no longer the dominant excitation mechanism of PNS oscillations \cite{MuMeLe25}, though accretion continues throughout the simulation. After this time, the results are qualitatively similar to the D15 model.

Finally, as all universal relations aside from Rodriguez et al. 2023 \cite{RoRaCh23} were derived using one- and two-dimensional CCSN simulations, we investigated each relation for two, two-dimensional \textsc{Chimera} models. In these models, we see the same qualitative behavior described for D15, until $\sim$600 ms after bounce. At this point, the ${}^2p_1$ mode relation of Rodriguez et al. 2023 \cite{RoRaCh23} as well as the ${}^2g_1$ mode relation of Torres-Forné et al. 2021 \cite{ToCeMu21} begin to deviate from the gfF significantly. We do not see the same deviation for the ${}^2f$ mode of Rodriguez et al. 2023 \cite{RoRaCh23} and the PNS mean density with the surface defined at $10^{11}$ g cm$^{-3}$. As these simulations run to later times, it is not immediately apparent if this deviation is due to a difference in simulation dimensionality or if it would appear at later times in the D15 model as well.

We highlight the discrepancies from the universal relations used in this paper and the gravitational wave frequencies computed directly from the CCSN simulation data not to disparage the individual relations themselves, but to emphasize the need for caution when applying any universal relation to a particular model. Currently, gravitational wave forms derived from CCSN simulations across multiple groups are being used to develop and inform the algorithms that will be used to detect and characterize the gravitational waves generated by the next galactic CCSN. We present here cases where universal relation predictions do not fit well with the simulation data. The origin of these discrepancies requires further study, but we can point to several known issues at the outset: 
\begin{enumerate}
    \item There is no unique definition for the boundary used in the linear perturbation analysis.
    \item In some cases, there is a discrepancy between the treatment of gravity in the simulations and the treatment of gravity in the linear perturbation analysis.
    \item Nearly all of the universal relations were developed using one- and two-dimensional models, not three-dimensional models.
    \item The sophistication of the models used to develop the universal relations varies considerably across the models.
\end{enumerate}
More concerning, a naive application of the universal relations can lead to potentially incorrect conclusions about the character of the gravitational wave emission, misidentifying either the mode generating the gravitational wave or the PNS property connected to the emission.

\funding{ 
A.M. acknowledges support from the National Science Foundation's Gravitational Physics Theory Program through grant PHY-2409148. P. M. is supported by the
National Science Foundation through its employee IR/D
program.

An award of computer time was provided by the Innovative and Novel Computational Impact on Theory and Experiment (INCITE) program. This research used resources of the Oak Ridge Leadership Computing Facility, which is a DOE Office of Science User Facility supported under contract DE-AC05-00OR22725.
This research used resources of the National Energy Research Scientific Computing Center (NERSC), a U.S. Department of Energy Office of Science User Facility located at Lawrence Berkeley National Laboratory, operated under contract No. DE-AC02-05CH11231}

\data{The data presented here for the D-series is part of the \textsc{Chimera} group's gravitational wave data release which can be found at \cite{OLCF_GW_M}.}

\printbibliography

\end{document}